\documentclass[11pt]{article}
\usepackage{nicefrac}
\usepackage[utf8]{inputenc}
\usepackage{amssymb}
\usepackage{amsmath}
\usepackage{color}

\usepackage{cite}

\usepackage[left=1.25in,right=1.25in,top=1.0in,bottom=2.10in]{geometry}

\def\SBigg#1{{\hbox{$\left#1\vbox to19.5\p@{}\right.\n@space$}}}

\newcommand*\laplace{\mathop{}\!\mathbin\bigtriangleup}

\allowdisplaybreaks 

\begin{document}

\title{The connection between Bohmian mechanics and many-particle quantum hydrodynamics} 
\author{Klaus Renziehausen$^*$
, Ingo Barth$^*$
        \\ \\
\textit{\small{Max Planck Institute of Microstructure Physics, Weinberg 2, 06120 Halle (Saale), Germany}} \\
\small{$^*$E-mail: ksrenzie@mpi-halle.mpg.de; barth@mpi-halle.mpg.de}}

\date{}
\maketitle

\begin{abstract} \noindent Bohm developed the Bohmian mechanics (BM), in which the Schrödinger equation is transformed into two differential equations: A continuity equation and an equation of motion similar to the Newtonian equation of motion. This transformation can be executed both for single-particle systems and for many-particle systems. Later, Kuzmenkov and Maksimov used basic quantum mechanics for the derivation of many-particle quantum hydrodynamics (MPQHD) including one differential equation for the mass balance and two differential equations for the momentum balance, and we extended their ana\-lysis in a prework [K.\ Renziehausen, I.\ Barth, Prog.\ Theor.\ Exp.\ Phys.\ {\bf 2018}, 013A05 (2018)] for the case that the particle ensemble consists of different particle sorts. The purpose of this paper is to show how the differential equations of MPQHD can be derived for such a particle ensemble with the differential equations of BM as a starting point. Moreover, our discussion clarifies that the differential equations of MPQHD are more suitable for an analysis of many-particle systems than the differential equations of BM because the differential equations of MPQHD depend on a single position vector only while the differential equations of BM depend on the complete set of all particle coordinates. 
\end{abstract}


\maketitle

\section{Introduction}
In 1952, Bohm used preworks of Madelung \cite{Madelung_1926_1,Madelung_1926_2} to develop his Bohmian mechanics (BM) \cite{Bohm_1952_1,Bohm_1952_2}. In BM, the momentum balance for a non-relativistic quantum mechanical system is described by an equation of motion, which is similar to the Newtonian equation of motion -- the only difference to the Newtonian equation of motion is that in the Bohmian equation of motion an additional quantity appears and is called the quantum potential. Moreover, one can derive a continuity equation for the conservation of the number of particles. \newline 
Bohm focused in his works \cite{Bohm_1952_1,Bohm_1952_2} mainly on systems with one particle or two particles. But still, he explained in \cite{Bohm_1952_1} an ansatz for applying his mechanics on many-particle systems. In this publication, we analyze a many-particle system and for this case, we call the above-mentioned equation, which is related to the conservation of the number of the particles, the many-particle continuity equation of BM. As can be deduced from a discussion in \cite{Wyatt_2005}, p.\ 8f., one finds a many-particle Bohmian equation of motion in the Lagrangian picture, the form of which is analogous to the equation of motion in the original publication \cite{Bohm_1952_1} of Bohm, and another equation, which is a version of this many-particle Bohmian equation of motion in the Eulerian picture. For brevity, we will frequently call these two equations the Lagrangian or the Eulerian version of the many-particle Bohmian equation of motion, respectively. \newline 
In addition, in 1999 Kuzmenkov and Maksimov derived for a non-relativistic many-particle quantum system the basic physics of many-particle quantum hydrodynamics (MPQHD) \cite{Kuzmenkov_1999}; this analysis included the derivation of a differential equation for the mass balance and two differential equations for the momentum balance. Due to the discussions in \cite{Renziehausen_2017} and the following analysis in this paper, we call the differential equation for the mass balance the 
continuity equation of MPQHD. In addition, we call the first of the two momentum balance equations the many-particle Ehrenfest equation of motion because its derivation can be related to the Ehrenfest theorem (\hspace{-0.03cm}\cite{Ehrenfest_1927} and \cite{Schwabl_2005}, p.\ 28ff.) -- and we call the second of the two mass balance equations the many-particle quantum Cauchy equation because of its analogy to the Cauchy equation of motion, which is well-known in classical hydrodynamics (\hspace{-0.18 cm} \cite{Stokes_1966} and \cite{Acheson_2005}, p.\ 205). As a contrast to the above-mentioned Bohmian differential equations, which depend on the complete set of all particle coordinates, the derivation of the differential equations of MPQHD includes an averaging over coordinates of all except one particle so that they only depend on the coordinates of a single position vector. \newline
Furthermore, Kuzmenkov and his coworkers developed MPQHD further to analyze spin effects  \cite{Kuzmenkov_2001_1,Andreev_2007} and Bose-Einstein-Condensates \cite{Andreev_2008}. In addition, applications of MPQHD were discussed for electrons in graphene \cite{Andreev_2012_a} and plasma effects \cite{Andreev_2012_b, Andreev_2013, Andreev_2014a, Andreev_2014b, Trukhanova_2013,Trukhanova_2015}. Hereby, in \cite{Andreev_2014a}, it is shortly mentioned how to apply MPQHD for systems where several sorts of particle are present, and in \cite{Andreev_2014a, Andreev_2014b, Trukhanova_2013,Trukhanova_2015}, the equations for MPQHD were discussed for the special case of two particle sorts in a plasma. Hereby, in \cite{Andreev_2014a, Andreev_2014b,Trukhanova_2015}, the MPQHD for electrons and a single ion sort were discussed, and in \cite{Trukhanova_2013}, these two sorts are electrons and positrons.  \newline 
Moreover, in \cite{Renziehausen_2017}, we developed the methods described in \cite{Kuzmenkov_1999,Andreev_2014a} further by deriving rigo\-rously the MPQHD equations for the general case that the particle ensemble includes several sorts of particle. As a result of this discussion, we found that there is both a many-particle continuity equation of MPQHD, a many-particle Ehrenfest equation of motion, and the many-particle quantum Cauchy equation for each individual particle sort and for the total particle ensemble, where all these differential equations only depend on one position vector, too. \newline   
In the following analysis, we again discuss a system of many particles of different sorts. Now, the purpose of this work is to take two differential equations related to BM as a starting point, namely the many-particle continuity equation of BM and the Eulerian version of the many-particle Bohmian equation of motion, and to find then these MPQHD differential equations as an ending point: We derive both for each individual particle sort and for the total particle ensemble the many-particle continuity equation of MPQHD and the many-particle quantum Cauchy equation. As intermediate results of this derivation, we will find the many-particle Ehrenfest equation of motion both for each individual particle sort and for the total particle ensemble, too. In addition, performing this calculation, we will have to do an averaging over the coordinates of all particles except one because -- as already stated --  the  differential equations of BM depend on the complete set of particle coordinates, while the differential equations of MPQHD, which we want to derive, only depend on a single position vector.  \newline 
By working out the details of this derivation, a gap is filled because it is known how the above-mentioned differential equations of BM and the above-mentioned differential equations of MPQHD can be found using basic quantum mechanics -- but to our knowledge, a derivation of these equations of MPQHD using equations of BM as a starting point has not been performed yet. Our motivation to do this derivation is that the reader can learn by following the analysis given here how on one hand these differential equations of BM, and on the other hand these differential equations of MPQHD are connected with each other. In addition, the following analysis in this publication shows that the differential equations of MPQHD are more suitable than the differential equations of BM for an analysis of many-particle systems because they only depend on one position vector while the differential equations of BM depend on the complete set of particle coordinates of the system. \newline
This advantage of the MPQHD ansatz over the BM ansatz becomes particularly clear for very large systems with more than hundred particles. Analoguously, for classical systems with so many particles, the application of a hydrodynamic ansatz is advantageous compared to solving Newtonian differential equations for large systems. \newline
For brevity, we will frequently omit the attribute ''many-particle'' for equations and quantities in the text below. However, the abbreviation MPQHD and the expression many-particle system is still used regularly. \newline 
In Sec.\ \ref{Basic Physics} of this paper, we recapitulate the application of BM for our system of many particles of different sorts, and we derive the continuity equation of BM and the Lagrangian and Eulerian versions of the Bohmian equation of motion starting from the Schrödinger equation. Then, in Sec.\ \ref{Connection}, the proof is given how we can use the continuity equation of BM and the Eulerian version of the Bohmian equation of motion as a starting point to derive the continuity equation of MPQHD and the quantum Cauchy equation both for each individual particle sort and for the total particle ensemble. Finally, in Sec.\ \ref{Summary}, a summary of the contents of this work is given.   
\section{Basic physics of BM for a many-particle system} \label{Basic Physics}
In this section, we refer to Bohm's approach \cite{Bohm_1952_1} to derive the continuity equation of BM and the Lagrangian and Eulerian versions of the Bohmian equation of motion -- besides the original publication of Bohm, we recommend to read the discussion about Bohmian mechanics for many-particle systems in \cite{Oriols_2012}, p.\ 60 ff., too. In analogy to our discussion in \cite{Renziehausen_2017}, here, we analyze a system with ${N_S}$ sorts of particle, where $\textnormal{A, B}$ stands for any number $\in \{1,2,\ldots, {N_S}\}$ which is related to one sort of particle. In addition, we call any $\textnormal{A}$th sort of particle also the sort of particle $\textnormal{A}$ or even shorter as the sort $\textnormal{A}$. There are indistinguishable $N(\textnormal{A})$ particles for each sort $\textnormal{A}$, and each of the $N(\textnormal{A})$ particles of the sort $\textnormal{A}$ has a mass $m_\textnormal{A}$. Moreover, we do not analyze spin effects in this publication (otherwise, we would have to regard that each particle of the sort $\textnormal{A}$ has a spin $s_\textnormal{A}$). In particular, all the following analysis is valid for the three cases that all particles are bosons, or that all particles are fermions, or that the particles of some sorts are fermions and the particles of the other sorts are bosons. \newline
As already mentioned in \cite{Renziehausen_2017}, it would be important for a many-particle system to discuss the property of the particles being either bosons or fermions if one uses a deconstruction of the wave function with the Hartree-Fock method as a sum over eigenfunctions in the occupation number space. In an analysis in \cite{Kuzmenkov_1999}, it was found that for such a deconstruction of the wave function a distinction between the cases that the particles are bosons or fermions is required. But we make neither in the calculations of our former work \cite{Renziehausen_2017} nor here such a deconstrucion of the wave function. As a consquence, all the calculations in this work are valid both for bosons and for fermions. \newline 
We designate the position vector of the $i$th particle of the sort $\textnormal{A}$ (so $i \in {1,2,\ldots, N(\textnormal{A})}$) as $\vec q_i^{\hspace{0.05 cm}\textnormal{A}}$ and name this particle the $(\textnormal{A},i)$ particle. Then, the complete set of particle coordinates $\vec Q$ is given by 
\begin{eqnarray}
\vec Q &=&  \left ( \vec q_1^{\hspace{0.05 cm}1}, \vec q_2^{\hspace{0.05 cm}1}, \ldots, \vec q_{N(1)}^{\hspace{0.05 cm}1}, \vec q_1^{\hspace{0.05 cm}2}, \ldots,  \vec q_{N(2)}^{\hspace{0.05 cm}2}, \ldots, \vec q_{1}^{\hspace{0.05 cm}N_S}, \ldots, \vec q_{N({N_S})}^{\hspace{0.05 cm}{N_S}} \right), \label{Particle Coordinate Set}
\end{eqnarray}
and the normalized total wave function of the system is $\Psi(\vec Q,t)$. This wave function obeys the Schrödinger equation 
\begin{eqnarray}
\mathrm{i} \hbar \; \frac{\partial \Psi (\vec Q,t)}{\partial t} &=& \hat H (\vec Q,t) \; \Psi (\vec Q,t). \label{Schroedinger equation}
\end{eqnarray}
In this equation the Hamiltonian appears as:
\begin{eqnarray}
\hat H (\vec Q,t) &=&  \sum_{\textnormal{A} = 1}^{N_S} \sum_{i = 1}^{N(\textnormal{A})} \frac {\left(\hat {\vec p}_i^{\hspace{0.05 cm} \textnormal{A}}\right)^2}{2 m_\textnormal{A}} + V(\vec Q,t), \label{Hamiltion operator with momentum operator}  
\end{eqnarray}
where $V(\vec Q,t)$ is a scalar potential, and $\hat {\vec p}_i^{\hspace{0.05 cm} \textnormal{A}}$ is the momentum operator of the  $(\textnormal{A},i)$ particle. This operator is given by 
\begin{eqnarray}
\hat {\vec p}_i^{\hspace{0.05 cm} \textnormal{A}} &=& \frac{\hbar}{\mathrm{i}} \nabla_i^\textnormal{A},
\end{eqnarray}
where $\nabla_i^\textnormal{A}$ is the nabla operator relative to the  $\vec q_i^{\hspace{0.05 cm}\textnormal{A}}$ coordinate. \newline
The Hamiltonian $\hat H(\vec Q,t)$ used in this work applies, e.\ g., to the interaction of molecules with external laser fields within the dipole approximation \cite{Bransden_2003}, p.\ 195 ff., or molecules without external fields. In the latter case, the potential is time-independent, so the Hamiltonian is time-independent, too. \newline 
Now, Bohm applied the following ansatz to transform the Schrödinger equation (\ref{Schroedinger equation}), in which the wave function is factorized as
\begin{eqnarray}
\Psi(\vec Q,t) = a(\vec Q,t) \exp \left[\frac{\mathrm{i} S(\vec Q,t)}{\hbar}\right],  \label{Bohm representation}
\end{eqnarray}
where $a(\vec Q,t)$ and $S(\vec Q,t)$ are real-valued functions. \newline 
As the next step, we insert the expression (\ref{Bohm representation}) for $\Psi(\vec Q,t)$ into the Schrödinger equation (\ref{Schroedinger equation}), then we multiply the resulting formula with $\exp [-\mathrm{i} S(\vec Q,t)/\hbar]$. After that, we separate the real and the imaginary parts of the outcome and find these two equations: 
\begin{align}
- a \frac{\partial S}{\partial t}   &= \sum_{\textnormal{A},i} \left[ \frac{1}{2 m_\textnormal{A}} a \left ( \nabla_i^\textnormal{A} S \right)^2 - \frac{\hbar^2}{2 m_\textnormal{A}} \laplace_i^\textnormal{A} a \right] + a V \label{Realteilgleichung}, \\
\hbar \frac{\partial a}{\partial t} &= - \hbar \sum_{\textnormal{A},i} \frac{1}{2 m_\textnormal{A}} \left[ a \laplace_i^\textnormal{A} S + 2 \left( \nabla_i^\textnormal{A} a \right) \left( \nabla_i^\textnormal{A} S \right) \right] \label{Imaginaerteilgleichung},
\end{align}
where $\laplace_i^\textnormal{A} = \nabla_i^\textnormal{A} \cdot \nabla_i^\textnormal{A}$ is the Laplace operator relative to $\vec q_i^{\hspace{0.05 cm}\textnormal{A}}$. \newline 
We initially concentrate on the transformation of Eqn.\ (\ref{Imaginaerteilgleichung}): We multiply this formula with $2a(\vec Q,t)/\hbar$, and after a straightforward calculation, we find: 
\begin{align}
\frac{\partial}{\partial t} a^2 &= - \sum_{\textnormal{A},i} \nabla_i^\textnormal{A} \left (a^2 \frac{\nabla_i^\textnormal{A} S}{m_\textnormal{A}} \right). \label{Vorstufe_zur_Kontinuitaetsgleichung} 
\end{align}
Then, we introduce the Bohmian particle density $D(\vec Q,t)$, the Bohmian velocity $\vec w_i^\textnormal{A}(\vec Q,t)$, and 
the Bohmian current density $\vec J_i^\textnormal{A}(\vec Q,t)$ of the $(\textnormal{A},i)$ particle:
\begin{align}
D(\vec Q ,t)                       &=  \left| \Psi(\vec Q,t) \right|^2 =a(\vec Q,t)^2,                 \label{Def_total_particle_density} \\
\vec w_i^\textnormal{A}(\vec Q,t)  &=  \frac{1}{m_\textnormal{A}} \nabla_i^\textnormal{A} S(\vec Q,t), \label{Def_velocity} \\
\vec J_i^\textnormal{A}(\vec Q,t)  &=  D(\vec Q ,t) \; \vec w_i^\textnormal{A}(\vec Q,t).              \label{Def_BM_current_density}
\end{align}
At this point, we make an excursion to the topic of the naming of some quantities which appear in our prework \cite{Renziehausen_2017} and in this work: \newline 
The different names of quantities used in \cite{Renziehausen_2017} and in this work are listed in Tab.\ \ref{Table1}. Our motivation to rename these quantities is that the Bohmian differential equations depend on the complete set of particle coordinates $\vec Q$, and the MPQHD differential equations depend on a single coordinate $\vec q$. So, in this work, we name the quantities according to Tab.\ \ref{Table1} Bohmian quantities if they depend on the complete set of particle coordinates $\vec Q$, or MPQHD quantities if they depend only on a single position vector $\vec q$, respectively. 
\begin{table} [t!]  
\begin{tabular}{|c|c|c|} \hline 
 {\small Symbol}                                & {\small Naming in \cite{Renziehausen_2017}}                      & {\small Naming in this work}                    \\ \hline \hline
 {\small $D(\vec Q,t)$}                         & {\small total particle density}                                  & {\small Bohmian particle density}               \\ \hline 
 {\small $\vec d_i^\textnormal{A}(\vec Q,t)$}   & {\small osmotic velocity}                                        & {\small Bohmian osmotic velocity}               \\                                                                                                                	  				       & {\small of the $(\textnormal{A},i)$ particle}                    & {\small of the $(\textnormal{A},i)$ particle}   \\ \hline  
 {\small $\vec w_i^\textnormal{A}(\vec Q,t)$}   & {\small velocity}                                                & {\small Bohmian velocity}                       \\
                                                & {\small of the $(\textnormal{A},i)$ particle}                    & {\small of the $(\textnormal{A},i)$ particle}   \\ \hline 
 {\small $\vec p_i^\textnormal{A}(\vec Q,t)$}   & {\small Was not used in \cite{Renziehausen_2017}}                                               & {\small Bohmian momentum}                       \\                                             &                                                                  & {\small of the $(\textnormal{A},i)$ particle}   \\ \hline   
 {\small $\vec J_i^\textnormal{A}(\vec Q,t)$}   & {\small Was not used in \cite{Renziehausen_2017}}                & {\small Bohmian current density}                \\ 
                                                &                                                                  & {\small of the $(\textnormal{A},i)$ particle}   \\ \hline 
 {\small $\rho_m^\textnormal{A}(\vec q,t)$}     & {\small one-particle mass density}                               & {\small MPQHD mass density}                       \\                                                                                                               
                                                & {\small for the particles of the sort $\textnormal{A}$}          & {\small for the particles of the sort $\textnormal{A}$} \\ \hline 
 {\small $\rho_m^\textnormal{tot}(\vec q,t)$}   & {\small total one-particle mass density}                         & {\small MPQHD total mass density}                 \\ \hline 
 {\small $\vec v^\textnormal{A}(\vec q,t)$}     & {\small mean particle velocity}                                  & {\small MPQHD particle velocity}                  \\                                                                                                  	                                       & {\small for the particles of the sort $\textnormal{A}$}          & {\small for the particles of the sort $\textnormal{A}$} \\ \hline 
 {\small $\vec v^{\hspace{0.5 mm} \textnormal{tot}}(\vec q,t)$}   & {\small mean particle velocity}                & {\small MPQHD particle velocity}                  \\
                                                & {\small for the total particle ensemble}                         & {\small for the total particle ensemble}        \\ \hline 
 {\small $\vec j_m^\textnormal{A}(\vec q,t)$}   & {\small mass current density}                                    & {\small MPQHD mass current density}               \\                
                                                & {\small for the particles of the sort $\textnormal{A}$}          & {\small for the particles of the sort $\textnormal{A}$} \\ \hline 
 {\small $\vec j_m^\textnormal{tot}(\vec q,t)$} & {\small total particle mass current density \cite{Renziehausen_2017_comment_A}}                     
                                                                                                                   & {\small MPQHD total mass current density} \\  \hline                                                                                                                                                                                                                                                                                                                                                                                                      
\end{tabular}
\caption{\noindent Overview over the naming of some quantities in our prework \cite{Renziehausen_2017} and in this work. The nomenclature used here facilitates the distinction between quantities related to BM and quantities related to MPQHD.} 
\label{Table1}
\end{table} 
 \newline \noindent Turning back to our derivations, using the definitions (\ref{Def_total_particle_density}), (\ref{Def_velocity}), and (\ref{Def_BM_current_density}), we can transform Eqn.\ (\ref{Vorstufe_zur_Kontinuitaetsgleichung}) into the continuity equation of BM for a many-particle system, which is related to the conservation of the number of particles. It is given by \cite{Bohm_1952_1}:
\begin{align}
\frac{\partial D(\vec Q,t)}{\partial t} &= - \sum_{\textnormal{A},i}  \nabla_i^\textnormal{A} \vec J_i^\textnormal{A}(\vec Q,t). \label{first_version_MPBCE} 
\end{align}
Now, we define the $\left[\sum_\textnormal{A} 3N(\textnormal{A}) \right]$-dimensional nabla operator $\nabla_{\vec Q}$ and the $\left[\sum_\textnormal{A} 3 N(\textnormal{A}) \right]$-dimensional Bohmian particle current density $\vec J(\vec Q,t)$:
\begin{align}
\nabla_{\vec Q} &= \left ( \nabla_1^{\hspace{0.05 cm}1}, \nabla_2^{\hspace{0.05 cm}1}, \ldots, \nabla_{N(1)}^{\hspace{0.05 cm}1}, \nabla_1^{\hspace{0.05 cm}2}, \ldots,  \nabla_{N(2)}^{\hspace{0.05 cm}2}, \ldots, \nabla_{1}^{\hspace{0.05 cm}N_S}, \ldots, \nabla_{N({N_S})}^{\hspace{0.05 cm}{N_S}} \right)^\textnormal{T}\hspace{-0.1 cm}, \label{Def_global_nabla} \\
\vec J(\vec Q,t)  &= \left ( \vec J_1^{\hspace{0.05 cm}1}, \vec J_2^{\hspace{0.05 cm}1}, \ldots, \vec J_{N(1)}^{\hspace{0.05 cm}1}, \vec J_1^{\hspace{0.05 cm}2}, \ldots,  \vec J_{N(2)}^{\hspace{0.05 cm}2}, \ldots, \vec J_{1}^{\hspace{0.05 cm}N_S}, \ldots, \vec J_{N({N_S})}^{\hspace{0.05 cm}{N_S}} \right)^\textnormal{T}\hspace{-0.1 cm}. \label{Def_global_w} 
\end{align}
Using these definitions, the continuity equation of BM can be written in a more compact form: 
\begin{align}
\frac{\partial D(\vec Q,t)}{\partial t} &= - \nabla_{\vec Q} \vec J(\vec Q,t). \label{second_version_MPBCE} 
\end{align}
Having found the continuity equation of BM, we derive the Bohmian equation of motion. For this task, we divide Eqn.\ (\ref{Realteilgleichung}) by $-a(\vec Q,t)$ and bring all terms on one side. Thus, we find: 
\begin{align}
\frac{\partial S}{\partial t} + \sum_{\textnormal A,i} \frac{\left(\nabla_i^\textnormal{A} S \right)^2}{2 m_\textnormal{A}} - \sum_{\textnormal A,i} \frac{\hbar^2}{2 m_\textnormal{A}} \frac{\laplace_i^\textnormal{A} a}{a} + V &= 0. \label{Vorstufe many-particle Hamilton-Jacobi equation}
\end{align}
Then, we define the quantum potential $V_{\textnormal{qu}}(\vec Q,t)$ as:
\begin{align}
V_{\textnormal{qu}}(\vec Q,t) &= -\sum_{\textnormal{A},i} \frac{\hbar^2}{4 m_\textnormal{A}} \left[ \frac{\laplace_i^\textnormal{A} D(\vec Q,t)}{D(\vec Q,t)} - \frac{\left(\nabla_i^\textnormal{A} D(\vec Q,t) \right)^2}{2 D(\vec Q,t)^2} \right], \label{quantum potential}                           
\end{align}
and one can show in a straightforward calculation by inserting the definition (\ref{Def_total_particle_density}) for the Bohmian particle density $D(\vec Q,t)$ into Eqn.\ (\ref{quantum potential}) that this equation for the quantum potential $V_{\textnormal{qu}}(\vec Q,t)$ can be rewritten as (\cite{Wyatt_2005}, p.\ 8 and \hspace{0.1mm}\cite{Kuzmenkov_1999}): 
\begin{align}
V_{\textnormal{qu}}(\vec Q,t) &= -\sum_{\textnormal{A},i} \frac{\hbar^2}{2 m_\textnormal{A}} \frac{\laplace_i^\textnormal{A} a(\vec Q,t)}{a(\vec Q,t)}. \label{quantum potential 2}                           
\end{align}
Then, we transform the formula (\ref{Vorstufe many-particle Hamilton-Jacobi equation}) by inserting Eqn.\ (\ref{quantum potential 2}) for the quantum potential  $V_{\textnormal{qu}}(\vec Q,t)$ and by inserting Eqn.\ (\ref{Def_velocity}) for the Bohmian velocity $\vec w_i^\textnormal{A}(\vec Q,t)$ of the $(\textnormal{A},i)$ particle. As an intermediate result, we find the quantum Hamilton-Jacobi equation \cite{Bohm_1952_1} (see also \cite{Wyatt_2005}, p.\ 8) for a many-particle system: 
\begin{align}
\frac{\partial S}{\partial t} + \sum_{\textnormal A,i} \frac{m_\textnormal{A} \left(\vec w_i^\textnormal{A} \right)^2}{2} + V_{\textnormal{qu}} + V &= 0. \label{Many-particle Hamilton-Jacobi equation}
\end{align}
As the next step, we apply the nabla operator $\nabla_{\vec Q}$ to this scalar equation and find the following $\left[\sum_\textnormal{A} 3N(\textnormal{A}) \right]$-dimensional equation: 
\begin{align}
\frac{\partial \left(\nabla_{\vec{Q}} S \right)}{\partial t} + \sum_{\textnormal A,i} \frac{m_\textnormal{A} \nabla_{\vec{Q}} \left[ \left( \vec w_i^\textnormal{A} \right)^2 \right]}{2} &= - \nabla_{\vec{Q}} \left(V_{\textnormal{qu}} + V \right).  \label{Nabla Many-particle Hamilton-Jacobi equation}
\end{align}
Now, we define the Bohmian momentum $\vec p_i^{\hspace{0.05 cm} \textnormal{A}}(\vec Q,t)$ of the $(\textnormal{A},i)$ particle and the $\left[\sum_\textnormal{A} 3N(\textnormal{A}) \right]$-dimensional Bohmian momentum $\vec p(\vec Q,t)$ as:
\begin{align}
\vec p_i^{\hspace{0.05 cm} \textnormal{A}}(\vec Q,t) &= m_\textnormal{A} \vec w_i^\textnormal{A} (\vec Q,t),  \label{Def_single_p}   \\
\vec p(\vec Q,t)                  &= \left (\vec p_1^{\hspace{0.05 cm}1}, \vec p_2^{\hspace{0.05 cm}1}, \ldots,  \vec p_{N(1)}^{\hspace{0.05 cm}1}, \vec p_1^{\hspace{0.05 cm}2}, \ldots, \vec p_{N(2)}^{\hspace{0.05 cm}2}, \ldots, \vec p_{1}^{\hspace{0.05 cm}N_S}, \ldots, \vec p_{N({N_S})}^{\hspace{0.05 cm}{N_S}} \right)^\textnormal{T}\hspace{-0.1 cm}. \label{Def_global_p} 
\end{align}
Having defined the Bohmian momentum $\vec p(\vec Q,t)$, we can write the first term on the left side of Eqn.\ (\ref{Nabla Many-particle Hamilton-Jacobi equation}) in the following manner: 
\begin{align}
\frac{\partial \left( \nabla_{\vec Q} S \right)}{\partial t} &= \frac{\partial \vec p}{\partial t}. \label{intermediate_result_partial_t} 
\end{align}
In the following calculations of this work, $x,y,$ and $z$ denote Cartesian vector indices or tensor indices, and sums using the Greek variables $\alpha$ or $\beta$ as sum variables are summing up over these indices $x,y,$ and $z$. \newline Then, for the vector component of the second term on the left side of Eqn.\ (\ref{Nabla Many-particle Hamilton-Jacobi equation}), which is related to the $x$ component of the $(\textnormal{B},1)$ particle, we find:
\begin{eqnarray}
\text{\small{$ \sum_{\textnormal{A},i} \frac{m_\textnormal{A}}{2} \frac{\partial}{\partial q_{1x}^{\textnormal{B}}}\left[ \left( w_{i x}^{\textnormal{A}} \right)^2 + \left( w_{i y}^{\textnormal{A}} \right)^2 + \left( w_{i z}^{\textnormal{A}} \right)^2 \right]$}} = \text{\small{$
 \sum_{\textnormal{A},i} m_\textnormal{A} \left( w_{i x}^{\textnormal{A}} \frac{\partial}{\partial q_{1x}^{\textnormal{B}}} w_{i x}^{\textnormal{A}}  + w_{i y}^{\textnormal{A}} \frac{\partial}{\partial q_{1x}^{\textnormal{B}}} w_{i y}^{\textnormal{A}} + w_{i z}^{\textnormal{A}} \frac{\partial}{\partial q_{1x}^{\textnormal{B}}} w_{i z}^{\textnormal{A}} \right)\hspace{-0.1 cm}.$}} \label{intermediate_result_derivation_of_square_1}
\end{eqnarray}
Moreover, using the definition (\ref{Def_velocity}) for the Bohmian velocity $\vec w_i^\textnormal{A}(\vec Q,t)$, we find that for any $\alpha \in \{x,y,z\}$ the following transformation is valid: 
\begin{align}
m_\textnormal{A} w_{i \alpha}^{\textnormal{A}} \frac{\partial}{\partial q_{1x}^{\textnormal{B}}}  w_{i \alpha}^{\textnormal{A}} &= w_{i \alpha}^{\textnormal{A}} \frac{\partial^2 S}{ \partial q_{1x}^{\textnormal{B}}  \partial q_{i \alpha}^{\textnormal{A}} }  \nonumber \\
&= m_\textnormal{B}  w_{i \alpha}^{\textnormal{A}} \frac{\partial}{\partial q_{i\alpha}^{\textnormal{A}}} w_{1 x}^{\textnormal{B}}. \label{intermediate_result_derivation_of_square_2}
\end{align}
As the next step, we insert Eqn.\ (\ref{intermediate_result_derivation_of_square_2}) into the intermediate result (\ref{intermediate_result_derivation_of_square_1}) and find:
\begin{align}
&  \sum_{\textnormal{A},i} m_\textnormal{A} \left( w_{i x}^{\textnormal{A}} \frac{\partial}{\partial q_{1x}^{\textnormal{B}}} w_{i x}^{\textnormal{A}}  + w_{i y}^{\textnormal{A}} \frac{\partial}{\partial q_{1x}^{\textnormal{B}}} w_{i y}^{\textnormal{A}} + w_{i z}^{\textnormal{A}} \frac{\partial}{\partial q_{1x}^{\textnormal{B}}} w_{i z}^{\textnormal{A}} \right) = \nonumber \\
= \; & \sum_{\textnormal{A},i} m_\textnormal{B} \left( w_{i x}^{\textnormal{A}} \frac{\partial}{\partial q_{ix}^{\textnormal{A}}} w_{1 x}^{\textnormal{B}}  + w_{i y}^{\textnormal{A}} \frac{\partial}{\partial q_{i y}^{\textnormal{A}}} w_{1 x}^{\textnormal{B}} + w_{i z}^{\textnormal{A}} \frac{\partial}{\partial q_{iz}^{\textnormal{A}}} w_{1 x}^{\textnormal{B}} \right) = \nonumber \\
= \; & \left ( \sum_{\textnormal{A},i} \sum_{\alpha} w_{i \alpha}^{\textnormal{A}} \frac{\partial}{\partial q_{i \alpha}^\textnormal{A}} \right) \underbrace{m_\textnormal{B} w_{1 x}^{\textnormal{B}}}_{= \; p_{1x}^{\textnormal{B}}} = \left [ \sum_{\textnormal{A},i} \left(\vec w_{i}^{\textnormal{A}} \nabla_i^\textnormal{A}\right) \right]  p_{1x}^{\textnormal{B}}. \label{intermediate_result_derivation_of_square_3}       
\end{align}
Thus, we found the result (\ref{intermediate_result_derivation_of_square_3}) for the vector component of the second term on the left side of Eqn.\ (\ref{Nabla Many-particle Hamilton-Jacobi equation}), which is related to the $x$ component of the $(\textnormal{B},1)$ particle. Therefore, the second term on the left side of Eqn.\ (\ref{Nabla Many-particle Hamilton-Jacobi equation}) can be written in this form: 
\begin{align}
\sum_{\textnormal{A},i} \frac{m_\textnormal{A}}{2} \nabla_{\vec Q} \left[ \left( \vec w_i^\textnormal{A} \right)^2 \right] &= \left [ \sum_{\textnormal{A},i} \left (\vec w_{i}^{\textnormal{A}} \nabla_i^\textnormal{A} \right) \right] \vec p. \label{intermediate_result_derivation_of_square_3b}  
\end{align}
Using Eqns.\ (\ref{intermediate_result_partial_t}) and (\ref{intermediate_result_derivation_of_square_3b}), the Eqn.\ (\ref{Nabla Many-particle Hamilton-Jacobi equation}) can be transformed into the following equation: 
\begin{align} 
\left [ \frac{\partial }{\partial t} +  \sum_{\textnormal{A},i} \left( \vec w_i^\textnormal{A} (\vec Q,t) \nabla_i^\textnormal{A} \right) \right] \vec p (\vec Q,t)  = - \nabla_{\vec Q} \left( V_{\textnormal{qu}}(\vec Q,t) + V(\vec Q,t) \right). \label{Eulerian_MPBEM}
\end{align}
Now, the Bohmian interpretation of the above-written equation describes the quantum dynamics of the particles in a manner that they move on trajectories. For example, the $(\textnormal{A},i)$ particle moves on a trajectory where it is at the time $t$ at the position $\vec q_i^{\hspace{0.05 cm}\textnormal{A}}(t)$. Thus, we can describe the trajectories of all the particles of the system together by a time-dependent set of particle coordinates $\vec Q(t)$ moving in a $[\sum_\textnormal{A} 3 N(\textnormal{A})]$-dimensional space on a single trajectory. That means we substitute the time-independent particle coordinates $\vec Q$ in Eqn.\ (\ref{Eulerian_MPBEM}) by time-dependent coordinates $\vec Q(t)$. \newline
In this description, we can interpret the left side of Eqn.\ (\ref{Eulerian_MPBEM}) as a total time derivative 
\begin{align}
\frac{\textnormal{d}\vec p(\vec Q(t),t)}{\textnormal{d}t} &= \frac{\partial \vec p (\vec Q(t),t)}{\partial t} +  \sum_{\textnormal{A},i} \left[ \vec w_i^\textnormal{A} (t) \; \nabla_i^\textnormal{A} \right] \vec p (\vec Q(t),t), \label{total_derivation_3}
\end{align}
where the Bohmian velocities are now given by $\vec w_i^\textnormal{A}(t) = \frac{\textnormal{d}\vec q_i^\textnormal{A}(t)}{\textnormal{d}t}$. \newline 
So, this total time derivative $\frac{\textnormal{d}\vec p(\vec Q(t),t)}{\textnormal{d}t}$ can be decomposed in two terms: Hereby, the first term $\frac{\partial \vec p(\vec Q(t),t)}{\partial t}$ is the local rate of change of the Bohmian momentum $\vec p(\vec Q(t),t)$, which is related to its explicit time-dependency. The second term $\sum_{\textnormal{A},i} \left[ \vec w_i^\textnormal{A}(\vec Q(t),t) \nabla_i^\textnormal{A} \right] \vec p(\vec Q(t),t)$ can be explained with the effect that the flow transports the fluid elements to other positions where the Bohmian momentum $\vec p(\vec Q(t),t)$ can differ. In addition, this second term is related to the implicit time dependency of the Bohmian momentum $\vec p(\vec Q(t),t)$ caused by the time-dependency of the set of particle coordinates $\vec Q(t)$. \newline 
As another consequence of the description of the movement of the particles within trajectories, we can substitute the functional dependencies of the Bohmian momentum $\vec p(\vec Q(t),t)$, namely the dependency on the set of particle coordinates $\vec Q(t)$, and the explicit dependency on the time $t$, by a single time dependency. \newline 
So, using the substitutions $\vec Q \rightarrow \vec Q(t)$, $\vec p(\vec Q(t),t) \rightarrow \vec p(t)$ and Eqn.\ (\ref{total_derivation_3}), we can transfer Eqn.\ (\ref{Eulerian_MPBEM}) into the Bohmian equation of motion for a many-particle system \cite{Bohm_1952_1,Bohm_1952_1_comment}:
\begin{align}
\frac{\textnormal{d} \vec p(t)}{\textnormal{d} t} &= - \nabla_{\vec Q(t)} \left[ V_\textnormal{qu}(\vec Q(t),t) + V(\vec Q(t),t) \right]. \label{Lagrangian_MPBEM}
\end{align}
This is an equation of motion similar to a Newtonian equation of motion having an ana\-logous form to the Bohmian equation of motion for a single particle published in the original work \cite{Bohm_1952_1} of Bohm. \newline 
Due to the derivation of the Bohmian equation of motion in \cite{Wyatt_2005}, p.\ 8f., one can interpret Eqn.\ (\ref{Eulerian_MPBEM}) as a version of the Bohmian equation of motion (\ref{Lagrangian_MPBEM}) in the Eulerian picture, where an observer at a fixed point in space watches the particles move by. \newline 
As a contrast, the Bohmian equation of motion (\ref{Lagrangian_MPBEM}) itself can be interpreted as an equation of motion in the Lagrangian picture, where the observer is follwing the set of particle coordinates $\vec Q(t)$ along its trajectory. In addition, we want to mention that,  if one solves (\ref{Lagrangian_MPBEM}) for a given right side of this equation and a given start Bohmian momentum $\vec p(t=t_0)$, then -- as discussed already in \cite{Bohm_1952_1} -- the given start momentum has to satisfy the constraint 
\begin{align}
\vec p(t_0) = \left[\nabla_{\vec Q(t)} S(\vec Q(t),t)\right]_{t=t_0}.
\end{align}
Following the discussion in \cite{Wyatt_2005}, p.\ 56f., this context can be related to a quantum version of Hamilton's principle. As an additional consequence of this quantum version of Hamilton's principle, the Bohmian velocities obey $\vec w_i^\textnormal{A}(t) = \frac{1}{m} \nabla_i^\textnormal{A} S(\vec Q(t),t)$, which is consistent with Eq.\ (\ref{Def_velocity}) after transforming this equation with the subtitutions $\vec Q \rightarrow \vec Q(t)$ and $\vec w_i^\textnormal{A}(\vec Q,t) \rightarrow  \vec w_i^\textnormal{A}(t)$ into an equation in the Lagrangian picture.   
\newline 
Now, we have found three equations being related to BM, namely the continuity equation of BM (\ref{second_version_MPBCE}), the Bohmian equation of motion (\ref{Lagrangian_MPBEM}) and the Eulerian version (\ref{Eulerian_MPBEM}) of the last-mentioned equation for a many-particle system. \newline
For our further analysis, we will use the continuity equation of BM (\ref{second_version_MPBCE}) and the Eulerian version (\ref{Eulerian_MPBEM}) of the Bohmian equation of motion.   
\section{Connection between BM and MPQHD} \label{Connection}
In this chapter, we will show that the two equations (\ref{second_version_MPBCE}) and (\ref{Eulerian_MPBEM}) being related to BM can be transformed in two quantum hydrodynamic equations: 
The continuity equation of BM (\ref{second_version_MPBCE}) can be transformed into the continuity equation of MPQHD and the Eulerian version of the Bohmian equation of motion (\ref{Eulerian_MPBEM}) can be transformed into the quantum Cauchy equation. Hereby, both the continuity equation of MPQHD and the quantum Cauchy equation only depend on a single position vector $\vec q$. As a contrast, the two equations (\ref{second_version_MPBCE}) and (\ref{Eulerian_MPBEM}) being related to BM depend on the complete set of particle coordinates $\vec Q$. \newline 
Please note that we derived the continuity equation of MPQHD and the quantum Cauchy equation already in \cite{Renziehausen_2017}, but in the derivations and discussions there, the connection between BM and MPQHD was not cleared -- this gap is filled in this work. Before we start with the derivations in this chapter, we point out these two details: \newline \newline
The first detail is that in \cite{Renziehausen_2017}, the continuity equation of MPQHD was named in another manner as the many-particle continuity equation because we had not to distinguish there between a Bohmian and a quantum hydrodynamic version of the continuity equation. \newline 
The second detail is that we presented in \cite{Renziehausen_2017} two different versions of the continuity equation of MPQHD and the quantum Cauchy equation -- namely, for both the  continuity equation of MPQHD and the quantum Cauchy equation there is one version for an individual particle sort $\textnormal{A}$ and another version for the total particle ensemble. We will show that the continuity equation of BM (\ref{second_version_MPBCE}) can be transformed into both versions of the continuity equation of MPQHD. Analogously, we will show that the Eulerian version of the Bohmian equation of motion (\ref{Eulerian_MPBEM}) can be transformed into both versions of the quantum Cauchy equation. \newline \newline 
Now, we will show the connection between BM and MPQHD. We start to carry out this task by transforming the continuity equation of BM (\ref{second_version_MPBCE}) into the continuity equation of MPQHD for the particle sort $\textnormal{A}$. \newline \newline Therefore, we multiply the continuity equation of BM (\ref{second_version_MPBCE}) by the factor $N(\textnormal{A}) m_\textnormal{A} \delta(\vec q - \vec q_1^{\hspace{0.05 cm} \textnormal{A}})$ and integrate it over the complete set of particle coordinates $\vec Q$. So, we find:
\begin{align}
N(\textnormal{A}) m_\textnormal{A} \hspace{-0.1 cm} \int  \hspace{-0.1 cm} d \vec Q \hspace{0.1 cm} \delta \hspace{-0.1 cm} \left(\vec q - \vec q_1^{\hspace{0.05 cm} \textnormal{A}}\right) \frac{\partial  D(\vec Q,t)}{\partial t} &= - N(\textnormal{A}) m_\textnormal{A} \hspace{-0.1 cm} \int \hspace{-0.1 cm} d \vec Q \hspace{0.1 cm}  \delta \hspace{-0.1 cm} \left(\vec q - \vec q_1^{\hspace{0.05 cm} \textnormal{A}}\right) \nabla_{\vec Q} \vec J(\vec Q,t). \label{Transformed_MPBCE}
\end{align}
Then, using the fact that the MPQHD mass density $\rho_m^\textnormal{A}(\vec q,t)$ for the particles of the sort $\textnormal{A}$ is given by \cite{Renziehausen_2017}
\begin{align}
\rho_m^\textnormal{A}(\vec q,t) &= N(\textnormal{A}) m_\textnormal{A} \; \int d \vec Q \; \delta \hspace{-0.1 cm} \left(\vec q - \vec q_1^{\hspace{0.05 cm} \textnormal{A}}\right)  D(\vec Q,t), 
\end{align}
we find for the left side of Eqn.\ (\ref{Transformed_MPBCE}):
\begin{align}
N(\textnormal{A}) m_\textnormal{A} \hspace{-0.1 cm} \int  \hspace{-0.1 cm} d \vec Q \hspace{0.1 cm} \delta \hspace{-0.1 cm} \left(\vec q - \vec q_1^{\hspace{0.05 cm} \textnormal{A}}\right) \frac{\partial  D(\vec Q,t)}{\partial t} &=  \frac{\partial}{\partial t} \left[ N(\textnormal{A}) m_\textnormal{A} \hspace{-0.1 cm}  \int \hspace{-0.1 cm} d \vec Q \hspace{0.1 cm}  \delta \hspace{-0.1 cm} \left(\vec q - \vec q_1^{\hspace{0.05 cm} \textnormal{A}}\right) D(\vec Q,t) \right] \nonumber \\
&= \frac{\partial \rho_m^\textnormal{A}(\vec q,t)}{\partial t}. \label{left_right_for_MPQHDCE}
\end{align}
For the transformation of the right side of Eqn.\ (\ref{Transformed_MPBCE}), we regard that 
\begin{align} 
\nabla_{\vec Q} \vec J(\vec Q,t) &=  \sum_{\textnormal{B},i} \nabla_i^\textnormal{B} \vec J_i^{\hspace{0.05 cm}\textnormal{B}}(\vec Q,t), 
\end{align}
so, it is given by: 
\begin{align}
- N(\textnormal{A}) m_\textnormal{A} \hspace{-0.1 cm} \int \hspace{-0.1 cm} d \vec Q \hspace{0.1 cm}  \delta \hspace{-0.1 cm} \left(\vec q - \vec q_1^{\hspace{0.05 cm} \textnormal{A}}\right) \nabla_{\vec Q}  \vec J(\vec Q,t) 
&= - N(\textnormal{A}) m_\textnormal{A}  \sum_{\textnormal{B},i}  \hspace{-0.1 cm} \int \hspace{-0.1 cm} d \vec Q \hspace{0.1 cm}  \delta \hspace{-0.1 cm} \left(\vec q - \vec q_1^{\hspace{0.05 cm} \textnormal{A}}\right) \nabla_i^\textnormal{B}  \vec J_i^{\hspace{0.05 cm}\textnormal{B}}(\vec Q,t). \label{right_side_transformed_MPBCE}
\end{align}
Using the volume element $\textnormal {d} \vec Q_{i}^{\textnormal{B}}$ for all coordinates except for the coordinate $\vec q_i^{\hspace{0.05 cm}\textnormal{B}}$, we can write the volume element $d \vec Q$ for the complete set of particles as 
\begin{align}
\textnormal {d} \vec Q &= \textnormal {d} \vec Q_{i}^{\textnormal{B}} \; \textnormal {d} \vec q_{i}^{\hspace{0.05 cm} \textnormal{B}}. \label{Product of Volume Elements}
\end{align}
As the next step, we transform Eqn.\ (\ref{right_side_transformed_MPBCE}) by splitting up the sum $\sum_{\textnormal{B},i}$ into the summand for the special case $\{ \textnormal{B} = \textnormal A, i = 1 \}$ and a sum over all the remaining summands. In addition, we use Eqn.\ (\ref{Product of Volume Elements}) for these remaining summands. Then, by applying the divergence theorem, we find that the integration over the coordinate $\vec q_i^{\hspace{0.05 cm} \textnormal{B}}$ for all these remaining summands can be transformed into an integral over the system boundary surface where the wave function vanishes. Thus, these remaining summands vanish, and only the summand for the special case $\{ \textnormal{B} = \textnormal A, i = 1 \}$ remains. By considering these ideas, we find: 
\begin{align}
& - N(\textnormal{A}) m_\textnormal{A}  \sum_{\textnormal{B},i}  \hspace{-0.1 cm} \int \hspace{-0.1 cm} d \vec Q \hspace{0.1 cm}  \delta \hspace{-0.1 cm} \left(\vec q - \vec q_1^{\hspace{0.05 cm} \textnormal{A}}\right) \nabla_i^\textnormal{B}  \vec J_i^{\hspace{0.05 cm}\textnormal{B}}(\vec Q,t) =  \nonumber \\
& \hspace{1 cm} = \; - N(\textnormal{A}) m_\textnormal{A} \hspace{-0.1 cm} \int \hspace{-0.1 cm} d \vec Q \hspace{0.1 cm}  \delta \hspace{-0.1 cm} \left(\vec q - \vec q_1^{\hspace{0.05 cm} \textnormal{A}}\right) \nabla_1^\textnormal{A} \vec J_1^\textnormal{A}(\vec Q,t) \nonumber  \\
& \hspace{1.5 cm} -  \underset{\{ \textnormal{B},j \} \neq \{  \textnormal{A}, 1 \}}{\sum_{\textnormal{B}=1}^{{N_S}} \sum_{i=1}^{N(\textnormal{B})}} 
  \int \hspace{-0.1 cm} d \vec Q_{i}^{\textnormal{B}} \hspace{0.1 cm}  \delta \hspace{-0.1 cm} \left(\vec q - \vec q_1^{\hspace{0.05 cm} \textnormal{A}}\right)  \hspace{-0.1 cm} \underbrace{\int \hspace{-0.1 cm} d \vec q_{i}^{\hspace{0.05cm} \textnormal{B}} \; \nabla_i^\textnormal{B} \vec J_i^{\hspace{0.05 cm}\textnormal{B}}(\vec Q,t)}_{= \; 0} = \nonumber \\
& \hspace{1 cm} = \; - N(\textnormal{A}) m_\textnormal{A} \hspace{-0.1 cm} \int \hspace{-0.1 cm} d \vec Q \hspace{0.1 cm}  \delta \hspace{-0.1 cm} \left(\vec q - \vec q_1^{\hspace{0.05 cm} \textnormal{A}}\right) \nabla_1^\textnormal{A}   \vec J_1^\textnormal{A}(\vec Q,t) \nonumber \\
& \hspace{1 cm} = \; - \nabla_{\vec q} \left[ N(\textnormal{A}) m_\textnormal{A} \hspace{-0.1 cm} \int \hspace{-0.1 cm} d \vec Q \hspace{0.1 cm}  \delta \hspace{-0.1 cm} \left(\vec q - \vec q_1^{\hspace{0.05 cm} \textnormal{A}}\right) \vec J_1^\textnormal{A}(\vec Q,t) \right].
\label{Preresult_right_for_MPQHDCE}
\end{align}
Now, the MPQHD mass current density $\vec j_m^\textnormal{A}(\vec q,t)$ for the particles of the sort $\textnormal{A}$ given by \cite{Renziehausen_2017,Renziehausen_2017_comment_B}
\begin{align}
\vec j_m^\textnormal{A}(\vec q,t) &=  N(\textnormal{A}) m_\textnormal{A}  \hspace{-0.1 cm} \int \hspace{-0.1 cm} d \vec Q \hspace{0.1 cm}  \delta \hspace{-0.1 cm} \left(\vec q - \vec q_1^{\hspace{0.05 cm} \textnormal{A}}\right)  \vec J_1^\textnormal{A}(\vec Q,t) \label{mass current density}
\end{align}
can be inserted in Eqn.\ (\ref{Preresult_right_for_MPQHDCE}), and we find the following result for the right side of Eqn.\ (\ref{Transformed_MPBCE}): 
\begin{align}
- N(\textnormal{A}) m_\textnormal{A}  \sum_{\textnormal{B},i}  \hspace{-0.1 cm} \int \hspace{-0.1 cm} d \vec Q \hspace{0.1 cm}  \delta \hspace{-0.1 cm} \left(\vec q - \vec q_1^{\hspace{0.05 cm} \textnormal{A}}\right) \nabla_i^\textnormal{B} \vec J_i^{\hspace{0.05cm} \textnormal{B}}  &= - \nabla_{\vec q} \hspace{0.05 cm} \vec  j_m^\textnormal{A}(\vec q,t). \label{result_right_for_MPQHDCE}
\end{align}
Finally, we can combine the results (\ref{left_right_for_MPQHDCE}) and (\ref{result_right_for_MPQHDCE}) for the left side and the right side of Eqn.\ (\ref{Transformed_MPBCE}) and get as a result the continuity equation of MPQHD for the particle sort $\textnormal{A}$ \cite{Renziehausen_2017}:
\begin{align}
\frac{\partial \rho_m^\textnormal{A}(\vec q,t)}{\partial t} &= - \nabla_{\vec q} \hspace{0.05 cm} \vec j_m^\textnormal{A}(\vec q,t). \label{MPQHDCE} 
\end{align}
Thus, we have proven that the continuity equation of MPQHD for the particle sort $\textnormal{A}$ (\ref{MPQHDCE}) can be derived from the continuity equation of BM (\ref{second_version_MPBCE}).
\newline \newline 
In addition, having completed this proof, it is trivial that the continuity equation of MPQHD for the total particle ensemble can be derived from the continuity equation of BM (\ref{second_version_MPBCE}), too, since in \cite{Renziehausen_2017}, it was already shown that the continuity equation of MPQHD for the total particle ensemble can be derived just by summing up the continuity equation of MPQHD for particles of a certain sort $\textnormal{A}$ over all sorts of particle. This continuity equation of MPQHD for the total particle ensemble has the form  
\begin{align}
\frac{\partial \rho_m^\textnormal{tot}(\vec q,t)}{\partial t} &= - \nabla_{\vec q} \hspace{0.05 cm}  \vec j_m^\textnormal{tot}(\vec q,t), \label{MPQHDCE_total} 
\end{align}
where the quantity $\rho_m^\textnormal{tot}(\vec q,t)$ is the MPQHD total mass density given by: 
\begin{align}
\rho_m^\textnormal{tot}(\vec q,t) &= \sum_{\textnormal{A}=1}^{N_S} \rho_m^\textnormal{A}(\vec q,t), \label{def_rho_tot}
\end{align}
and the quantity $\vec j_m^\textnormal{tot}(\vec q,t)$ is the MPQHD total mass current density given by: 
\begin{align}
\vec j_m^\textnormal{tot}(\vec q,t) &= \sum_{\textnormal{A}=1}^{N_S} \vec j_m^\textnormal{A}(\vec q,t).  \label{def_j_tot}
\end{align}
So, we have shown that one can take the continuity equation of BM (\ref{second_version_MPBCE}) as a starting point to derive both the continuity equation of MPQHD for the particle sort $\textnormal{A}$ (\ref{MPQHDCE}) and the continuity equation of MPQHD for the total particle ensemble (\ref{MPQHDCE_total}). \newline \newline 
In the following calculation, we will take the Eulerian version of the Bohmian equation of motion (\ref{Eulerian_MPBEM}) as a starting point to derive both the quantum Cauchy equation for the particle sort $\textnormal{A}$ and the quantum Cauchy equation for the total particle ensemble. \newline 
We start this calculation by taking into account the three vector components of the Eulerian version of the Bohmian equation of motion (\ref{Eulerian_MPBEM}), which are related to the $(\textnormal{A},1)$ particle. 
For these three vector components, the following differential equation holds: 
\begin{align} 
m_{\textnormal A} \left[ \frac{\partial}{\partial t} + \sum_{\textnormal{B},i} \left( \vec w_i^\textnormal{B}(\vec Q,t) \nabla_i^\textnormal{B} \right) \right] \vec w_i^\textnormal{A}(\vec Q,t)  &= - \nabla_1^\textnormal{A} \left [ V_{\textnormal{qu}}(\vec Q,t) + V(\vec Q,t) \right]. \label{Total_derivation_2}
\end{align}
Now, we multiply Eqn.\ (\ref{Total_derivation_2}) by $N(\textnormal{A}) D(\vec Q,t) \delta(\vec q - \vec q_1^{\hspace{0.05 cm} \textnormal{A}})$ and integrate it over the complete set of coordinates $\vec Q$. So, we find: 
\begin{align}
N(\textnormal{A}) m_\textnormal{A} \hspace{-0.1 cm} \int \hspace{-0.1 cm} d \vec Q \hspace{0.1 cm} \delta(\vec q - \vec q_1^{\hspace{0.05 cm} \textnormal{A}} )  \hspace{0.1 cm}  D   \left[ \frac{\partial}{\partial t} + \sum_{\textnormal{B},i} \left( \vec w_i^\textnormal{B}(\vec Q,t) \nabla_i^\textnormal{B} \right) \right] \hspace{-0.1 cm}  \vec w_1^\textnormal{A}(\vec Q,t) & = \nonumber \\ 
 - N(\textnormal{A})  \hspace{-0.1 cm} \int \hspace{-0.1 cm} d \vec Q \hspace{0.1 cm} \delta(\vec q - \vec q_1^{\hspace{0.05 cm} \textnormal{A}}) & \hspace{0.1 cm} D \hspace{0.1 cm} \nabla_1^\textnormal{A} \left [ V_{\textnormal{qu}} + V \right]. \label{MPBEM_transformed}
\end{align}
As the next step, we introduce the force density $\vec f^\textnormal{A}(\vec q,t)$ \cite{Kuzmenkov_1999,Renziehausen_2017,Renziehausen_2017_comment_C} and the quantum force density $\vec f^\textnormal{A}_{\textnormal{qu}}(\vec q,t)$ for the particles of the sort $\textnormal{A}$ (\hspace{0.1mm}\cite{Kuzmenkov_1999,Kuzmenkov_1999_comment} and cf.\ \cite{Wyatt_2005}, p. 326): 
\begin{align}
\vec f^\textnormal{A}(\vec q,t) &= N(\textnormal{A}) \hspace{-0.1 cm} \int \hspace{-0.1 cm} d \vec Q \hspace{0.1 cm} \delta(\vec q - \vec q_1^{\hspace{0.05 cm} \textnormal{A}} )  \hspace{0.1 cm} D(\vec Q,t) \left[ - \nabla_1^\textnormal{A} V(\vec Q,t) \right],  \label{force_density} \\
\vec f^\textnormal{A}_{\textnormal{qu}}(\vec q,t) &= N(\textnormal{A}) \hspace{-0.1 cm} \int \hspace{-0.1 cm} d \vec Q \hspace{0.1 cm} \delta(\vec q - \vec q_1^{\hspace{0.05 cm} \textnormal{A}} )  \hspace{0.1 cm} D(\vec Q,t) \left[ - \nabla_1^\textnormal{A} V_{\textnormal{qu}}(\vec Q,t) \right]. \label{quantum_force_density}
\end{align}
Using Eqns.\ (\ref{force_density}) and (\ref{quantum_force_density}), we can rewrite the right side of Eqn.\ (\ref{MPBEM_transformed}) in the following form: 
\begin{align}
- N(\textnormal{A})  \hspace{-0.1 cm} \int \hspace{-0.1 cm} d \vec Q \hspace{0.1 cm} \delta(\vec q - \vec q_1^{\hspace{0.05 cm} \textnormal{A}})  \hspace{0.1 cm}  D(\vec Q,t)  \hspace{0.1 cm} \nabla_1^\textnormal{A} \left [ V_{\textnormal{qu}} (\vec Q,t) + V (\vec Q,t) \right] &= \vec f^\textnormal{A}_{\textnormal{qu}}(\vec q,t) + \vec f^\textnormal{A}(\vec q,t). \label{MPBEM_transformed_rewritten} 
\end{align}
As an intermediate step, we introduce the momentum flow density tensor $\underline{\underline{\Pi}}^\textnormal{A}(\vec q,t)$ for the sort of particle $\textnormal{A}$. \newline 
As we discussed already in \cite{Renziehausen_2017}, there are several versions of this tensor $\underline{\underline{\Pi}}^\textnormal{A}(\vec q,t)$ because it is not this tensor itself which is physically relevant but only its tensor divergence $\nabla \underline{\underline{\Pi}}^\textnormal{A}(\vec q,t)$. Moreover, we pointed out in the mentioned reference that we recommend to use the Wyatt version of this tensor, and therefore, in this work, we will only use this tensor version. The naming of this tensor version is motivated by the form of the formula (1.57) in R. E. Wyatts book \cite{Wyatt_2005}, p.\ 31, and it is given by (hereby, $\underline{\underline{1}}$ is the unit matrix, and the operation symbol $\otimes$ interrelates two vectors to a dyadic product) \cite{Renziehausen_2017}:
\begin{align}
\underline{\underline{\Pi}}^{\textnormal{A}}(\vec q,t) &= \underline{\underline{1}} P_\textnormal{A} + N(\textnormal{A}) m_{\textnormal{A}}  \int \textnormal{d} \vec Q \; \delta (\vec q - \vec q_1^{\hspace{0.05 cm} \textnormal{A} } ) \; D \; \hspace{-0.1cm} \left[ \left(\vec w_{1}^\textnormal{A} \otimes \vec w_{1}^\textnormal{A} \right)  + \left(\vec d_{1}^{\hspace{0.05 cm} \textnormal{A}}  \otimes \vec d_{1}^{\hspace{0.05 cm} \textnormal{A}}  \right) \right] \hspace{-0.1cm}. \label{momentum_flow_densitytensor} 
\end{align}
Therefore, its Cartesian tensor elements are given by: 
\begin{align}
\Pi_{\alpha \beta}^{\textnormal{A}}(\vec q,t) &=  P_\textnormal{A} \delta_{\alpha \beta} + N(\textnormal{A}) m_\textnormal{A} \int \textnormal{d} \vec Q \; \delta (\vec q - \vec q_1^{\hspace{0.05 cm} \textnormal{A} } ) \; D \; \left (w_{1 \alpha}^\textnormal{A} w_{1 \beta}^\textnormal{A} + d_{1 \alpha}^\textnormal{A} d_{1 \beta}^\textnormal{A} \right) \hspace{-0.1cm}. \label{momentum_flow_densitytensorelements} 
\end{align}
The quantity $P_\textnormal{A}(\vec q,t)$, which appears in Eqns.\ (\ref{momentum_flow_densitytensor}) and (\ref{momentum_flow_densitytensorelements}), is the scalar quantum pressure for the particles of the sort $\textnormal{A}$, and this quantity is given by (\hspace{-0.1mm}\cite{Renziehausen_2017} and cf.\ \cite{Wyatt_2005}, p.\ 326):
\begin{align}
P_\textnormal{A}(\vec q,t) &=  - N(\textnormal{A}) \frac{\hbar^2}{4 m_\textnormal{A}} \int \textnormal{d} \vec Q \; \delta (\vec q - \vec q_1^{\hspace{0.05 cm}\textnormal{A}} ) \laplace_1^\textnormal{A} D. \label{scalar_pressure}
\end{align}
In addition, in Eqn.\ (\ref{momentum_flow_densitytensor}), the dyadic product of a vector $\vec d_i^{\hspace{0.05 cm}\textnormal{A}}(\vec Q,t)$ for the case $i=1$ appears; this vector is defined by 
\begin{align}
\vec d_i^{\hspace{0.05 cm} \textnormal{A}}(\vec Q,t) &= - \frac{\hbar}{2 m_\textnormal{A}} \frac{\nabla_i^\textnormal{A} {D}}{D}. \label{definition d}
\end{align}
The quantity $\vec d_i^{\hspace{0.05 cm} \textnormal{A}}(\vec Q,t)$ is named Bohmian osmotic velocity of the $(\textnormal{A},i)$ particle cor\-res\-pon\-ding to the nomenclature in  \cite{Wyatt_2005}, p.\ 327 and \hspace{-0.1mm}\cite{Renziehausen_2017}. The Bohmian osmotic velocity  $\vec d_i^{\hspace{0.05 cm} \textnormal{A}}(\vec Q,t)$ is the quantum analog to the Bohmian particle velocity $\vec w_i^\textnormal{A}(\vec Q,t)$, and in addition, this quantity is related to the shape of $D(\vec Q,t)$. \newline
Moreover, in \cite{Renziehausen_2017}, we discussed that one can split the momentum flow density tensor $\underline{\underline{\Pi}}^\textnormal{A}(\vec q,t)$ for the particles of the sort $\textnormal{A}$ in a classcial part and a quantum part: 
\begin{align}
\underline{\underline{\Pi}}^\textnormal{A}(\vec q,t) &= \underline{\underline{\Pi}}^\textnormal{A,cl}(\vec q,t) + \underline{\underline{\Pi}}^\textnormal{A,qu}(\vec q,t). \label{split_tensor}
\end{align}
Hereby, the classical part is given by
\begin{align}
\underline{\underline{\Pi}}^\textnormal{A,cl}(\vec q,t) &=  N(\textnormal{A}) m_{\textnormal{A}} \int \textnormal{d} \vec Q \; \delta (\vec q - \vec q_1^{\hspace{0.05 cm} \textnormal{A} }) \; D \; \hspace{-0.1cm} \left(\vec w_{1}^\textnormal{A} \otimes \vec w_{1}^\textnormal{A} \right) \hspace{-0.1cm}, \label{momentum_flow_densitytensor no N_k tensor version} 
\end{align}
and its matrix elements are
\begin{align}
\Pi^\textnormal{A,cl}_{\alpha \beta}(\vec q,t) &=  N(\textnormal{A}) m_{\textnormal{A}} \int \textnormal{d} \vec Q \; \delta (\vec q - \vec q_1^{\hspace{0.05 cm} \textnormal{A} }) \; D \;  w_{1\alpha}^\textnormal{A}  w_{1\beta}^\textnormal{A}. \label{classic_momentum_flow_densitytensor} 
\end{align}
And the quantum part is given by 
\begin{align}
\underline{\underline{\Pi}}^\textnormal{A,qu}(\vec q,t) &= \underline{\underline{1}} P_\textnormal{A} + N(\textnormal{A}) m_{\textnormal{A}} \int \textnormal{d} \vec Q \; \delta (\vec q - \vec q_1^{\hspace{0.05 cm} \textnormal{A} }) \; D \; \hspace{-0.1cm} \left( \vec d_{1}^{\hspace{0.05cm}\textnormal{A}} \otimes \vec d_{1}^{\hspace{0.05cm}\textnormal{A}} \right) \label{quantum_momentum_flow_densitytensor} 
\end{align}
with according matrix elements 
\begin{align}
\Pi^\textnormal{A,qu}_{\alpha \beta}(\vec q,t) &=  P_\textnormal{A} \delta_{\alpha \beta} + N(\textnormal{A}) m_{\textnormal{A}} \int \textnormal{d} \vec Q \; \delta (\vec q - \vec q_1^{\hspace{0.05 cm} \textnormal{A} }) \; D \; d_{1\alpha}^\textnormal{A}  d_{1\beta}^\textnormal{A}. \label{quantum_momentum_flow_densitytensor_elements} 
\end{align}
Now, we will prove that the quantum force $\vec f_{\textnormal{qu}}^\textnormal{A}(\vec q,t)$ is equal to the negative tensor divergence of the quantum part $\underline{\underline{\Pi}}^{\textnormal{A,qu}}(\vec q,t)$ of the momentum flow density tensor. So, we have to show that the following equation  holds, where $\vec e_\beta$, $\beta \in \{x,y,z\}$ are the Cartesian basis vectors: 
\begin{align}
\vec f_{\textnormal{qu}}^{\textnormal{A}}(\vec q,t) &= - \nabla_{\vec q} \hspace{0.05cm} \underline{\underline{\Pi}}^\textnormal{A,qu}(\vec q,t) \equiv - \sum_{\alpha,\beta} \frac{\partial \Pi_{\alpha \beta}^\textnormal{A,qu}(\vec q,t)}{\partial q_{\alpha}} \vec e_\beta. \label{Connection_quantum_force_force}
\end{align}
Therefore, we transform Eqn.\ (\ref{quantum_force_density}) for the quantum force density $\vec f^\textnormal{A}_{\textnormal{qu}}(\vec q,t)$ using Eqn.\ (\ref{quantum potential}) for the quantum potential $V_{\textnormal{qu}}(\vec q,t)$: 
\begin{align}
\vec f^\textnormal{A}_{\textnormal{qu}}(\vec q,t) =& \; N(\textnormal{A}) \hspace{-0.1 cm} \int \hspace{-0.1 cm} d \vec Q \hspace{0.1 cm} \delta(\vec q - \vec q_1^{\hspace{0.05 cm} \textnormal{A}} )  \hspace{0.1 cm} D \left[ - \nabla_1^\textnormal{A} V_{\textnormal{qu}} \right] \nonumber \\
=& -N(\textnormal{A}) \hspace{-0.1 cm}  \int \hspace{-0.1 cm}  d \vec Q \hspace{0.1 cm} \delta(\vec q - \vec q_1^{\hspace{0.05 cm} \textnormal{A}})  \hspace{0.1 cm} \nabla_1^\textnormal{A} \left ( D \hspace{0.05 cm}  V_{\textnormal{qu}} \right) +  N(\textnormal{A}) \hspace{-0.1 cm}   \int  \hspace{-0.1 cm}  d \vec Q \hspace{0.1 cm} \delta(\vec q - \vec q_1^{\hspace{0.05 cm} \textnormal{A}})  \hspace{0.1 cm}  V_{\textnormal{qu}} \hspace{0.05 cm} \nabla_1^\textnormal{A}  D   \nonumber \\
=& -N(\textnormal{A}) \hspace{-0.1 cm}  \int \hspace{-0.1 cm}  d \vec Q \hspace{0.1 cm} \delta(\vec q - \vec q_1^{\hspace{0.05 cm} \textnormal{A}})  \hspace{0.1 cm} \nabla_1^\textnormal{A} \left \lbrace -\sum_{\textnormal{B},i} \frac{\hbar^2}{4 m_\textnormal{B}} \left [ \laplace_i^\textnormal{B} D - \frac{\left(\nabla_i^\textnormal{B} D \right)^2}{2 D} \right] \right \rbrace \nonumber \\
&- N(\textnormal{A}) \hspace{-0.1 cm} \int \hspace{-0.1 cm} d \vec Q \hspace{0.1 cm} \delta(\vec q - \vec q_1^{\hspace{0.05 cm} \textnormal{A}})  \hspace{0.1 cm} \sum_{\textnormal{B},i} \frac{\hbar^2}{4 m_\textnormal{B}} \left [ \frac{\laplace_i^\textnormal{B} D}{D} - \frac{\left(\nabla_i^\textnormal{B} D \right)^2}{2 D^2} \right]  \nabla_1^\textnormal{A} D  \nonumber \\
=& -N(\textnormal{A}) \hspace{-0.1 cm} \int \hspace{-0.1 cm}  d \vec Q \hspace{0.1 cm} \delta(\vec q - \vec q_1^{\hspace{0.05 cm} \textnormal{A}})  \hspace{0.1 cm}
\nabla_1^\textnormal{A} \left[- \sum_{\textnormal{B},i} \frac{\hbar^2}{4 m_\textnormal{B}} \laplace_i^\textnormal{B} D \right] \nonumber \\
&   -N(\textnormal{A}) \hspace{-0.1 cm} \int \hspace{-0.1 cm}  d \vec Q \hspace{0.1 cm} \delta(\vec q - \vec q_1^{\hspace{0.05 cm} \textnormal{A}})  \hspace{0.1 cm}
   \sum_{\textnormal{B},i} \frac{\hbar^2}{4 m_\textnormal{B}} \; \times \nonumber  \\  
&  \; \; \times \left \lbrace
  \frac{ \left( \laplace_i^\textnormal{B} D \right) \left ( \nabla_1^\textnormal{A} D \right ) }{D}    
- \frac{ \left( \nabla_i^\textnormal{B} D \right)^2 \left ( \nabla_1^\textnormal{A} D \right ) }{2 D^2} 
+ \nabla_1^\textnormal{A} \left[ \frac{ \left (\nabla_i^\textnormal{B} D \right)^2 }{2 D} \right]     
   \right \rbrace. \label{intermediate result for quantum force}    
\end{align}
Now, we transform the term appearing in the first line of Eqn.\ (\ref{intermediate result for quantum force}) by regarding in a argumentation similar to the transformation of Eqn.\ (\ref{right_side_transformed_MPBCE}) to Eqn.\ (\ref{Preresult_right_for_MPQHDCE}) that only the summand for $\{\textnormal{B}=\textnormal{A}, i=1\}$ does not vanish for this term. In addition, we use the definition (\ref{scalar_pressure}) for the scalar pressure $P_\textnormal{A}(\vec q,t)$ and find: \newpage \noindent
\begin{align}
& - N(\textnormal{A}) \hspace{-0.1 cm} \int \hspace{-0.1 cm}  d \vec Q \hspace{0.1 cm} \delta(\vec q - \vec q_1^{\hspace{0.05 cm} \textnormal{A}})  \hspace{0.1 cm}
\nabla_1^\textnormal{A} \left[- \sum_{\textnormal{B},i} \frac{\hbar^2}{4 m_\textnormal{B}} \laplace_i^\textnormal{B} D \right] = \nonumber \\
= & - N(\textnormal{A}) \hspace{-0.1 cm} \int \hspace{-0.1 cm}  d \vec Q \hspace{0.1 cm} \delta(\vec q - \vec q_1^{\hspace{0.05 cm} \textnormal{A}})  \hspace{0.1 cm}
\nabla_1^\textnormal{A} \left[- \frac{\hbar^2}{4 m_\textnormal{A}} \laplace_1^\textnormal{A} D \right] \nonumber \\
= & - \nabla_{\vec q} \left[ - N(\textnormal{A}) \frac{\hbar^2}{4 m_\textnormal{A}} \hspace{-0.1 cm} \int \hspace{-0.1 cm}  d \vec Q \hspace{0.1 cm} \delta(\vec q - \vec q_1^{\hspace{0.05 cm} \textnormal{A}}) \hspace{0.1 cm} \laplace_1^\textnormal{A} D \right] \nonumber \\
= & - \nabla_{\vec q} \hspace{0.05 cm} P_\textnormal{A} \; = \; -  \sum_\alpha \frac{\partial}{\partial q_\alpha} \left( \delta_{\alpha \beta} \hspace{0.05 cm} P_\textnormal{A} \right) \; = \; - \nabla_{\vec q} \left( \hspace{0.05 cm} \underline{\underline{1}} \hspace{0.05 cm}  P_\textnormal{A} \right). \label{intermediate result for quantum force 2}
\end{align}
Note that in the last line of the calculation above, we rewrote the gradient of the scalar quantity $P_\textnormal{A}(\vec q,t)$ into the divergence of the tensor  $\left[ \hspace{0.05 cm} \underline{\underline{1}} \hspace{0.05 cm}  P_\textnormal{A}(\vec q,t) \right]$. \newline
Our next step is to transform the term in curly brackets in Eqn.\ (\ref{intermediate result for quantum force}). 
Using the abbreviation $\frac{\partial}{\partial q_{i \alpha}^\textnormal{A}} = \partial_{i \alpha}^\textnormal{A}$ and the definition (\ref{definition d}) of the Bohmian osmotic velocity $\vec d_i^{\hspace{0.05 cm}\textnormal{A}}(\vec Q,t)$, we find in a straightforward calculation the following result for the $x$ component of the term in curly brackets in Eqn.\  (\ref{intermediate result for quantum force}): 
\begin{align}
&  \left \lbrace
  \frac{ \left( \laplace_i^\textnormal{B} D \right) \left ( \nabla_1^\textnormal{A} D \right ) }{D}    
- \frac{ \left( \nabla_i^\textnormal{B} D \right)^2 \left ( \nabla_1^\textnormal{A} D \right ) }{2 D^2} 
+ \nabla_1^\textnormal{A} \left[ \frac{ \left (\nabla_i^\textnormal{B} D \right)^2 }{2 D} \right]     
   \right \rbrace_x = \nonumber \\ 
=\; & \sum_\alpha \left \lbrace \frac{ \left( \partial_{i \alpha}^\textnormal{B} \partial_{i \alpha}^\textnormal{B} D \right) \left( \partial_{1 x}^\textnormal{A} D \right)}{D} 
- \frac{\left( \partial_{i \alpha}^\textnormal{B} D \right)^2 \left(\partial_{1 x}^\textnormal{A} D \right)}{2 D^2}
+ \hspace{0.1 cm} \partial_{1 x}^\textnormal{A} \left[ \frac{ \left( \partial_{i \alpha}^\textnormal{B} D \right)^2}{2 D} \right] \right \rbrace = \nonumber \\
=\; & \sum_\alpha  \left \lbrace \partial_{i \alpha}^\textnormal{B} \left[ \frac{ \left( \partial_{i \alpha}^\textnormal{B} D \right) \left( \partial_{1 x}^\textnormal{A} D \right)}{D} \right]   
- \frac{\left( \partial_{i \alpha}^\textnormal{B} D \right)  \left( \partial_{i \alpha}^\textnormal{B} \partial_{1 x}^\textnormal{A} D  \right)}{D}
+ \frac{\left( \partial_{i \alpha}^\textnormal{B} D \right)^2 \left(\partial_{1 x}^\textnormal{A} D \right)}{D^2} \right. \nonumber \\
\; & \left. \; \; \; \; \; \; \;  - \; \frac{\left( \partial_{i \alpha}^\textnormal{B} D \right)^2 \left(\partial_{1 x}^\textnormal{A} D \right)}{2 D^2} 
+ \frac{\left( \partial_{i \alpha}^\textnormal{B} D \right) \left(\partial_{1 x}^\textnormal{A} \partial_{i \alpha}^\textnormal{B} D \right)}{D}
- \frac{\left( \partial_{i \alpha}^\textnormal{B} D \right)^2 \left(\partial_{1 x}^\textnormal{A} D \right)}{2 D^2} 
\nonumber \right \rbrace \\
= \;& \sum_\alpha \partial_{i \alpha}^\textnormal{B} \left [ \frac{ \left( \partial_{i \alpha}^\textnormal{B} D \right) \left( \partial_{1 x}^\textnormal{A} D \right)}{D} \right ] \nonumber \\
= \;& \frac{4 m_\textnormal{A}  m_\textnormal{B}}{\hbar^2} \sum_\alpha \partial_{i \alpha}^\textnormal{B} \left( D \; d_{i \alpha}^\textnormal{B} \; d_{1 x}^\textnormal{A} \right). \label{intermediate result for quantum force 3}
\end{align}
So, using a dyadic product $ \vec d_{i}^{\hspace{0.07 cm}\textnormal{B}} \otimes \vec d_{1}^{\hspace{0.05cm} \textnormal{A}}$, we can write the term in curly brackets in Eqn.\  (\ref{intermediate result for quantum force}) as:
\begin{align}
\hspace{-0.15 cm} \frac{ \left( \laplace_i^\textnormal{B} D \right) \left ( \nabla_1^\textnormal{A} D \right ) }{D}    
- \frac{ \left( \nabla_i^\textnormal{B} D \right)^2 \left ( \nabla_1^\textnormal{A} D \right ) }{2 D^2} 
+ \nabla_1^\textnormal{A} \left[ \frac{ \left (\nabla_i^\textnormal{B} D \right)^2 }{2 D} \right]   =& \;  \frac{4 m_\textnormal{A} m_\textnormal{B}}{\hbar^2} \nabla_{i}^\textnormal{B} \left[ D  \left( \vec d_{i}^{\hspace{0.07 cm}\textnormal{B}} \otimes \vec d_{1}^{\hspace{0.05cm} \textnormal{A}} \right) \right] \hspace{-0.07 cm}. \label{intermediate result for quantum force 4}
\end{align} \newpage
As the next step, we transform Eqn.\ (\ref{intermediate result for quantum force}) using the intermediate results (\ref{intermediate result for quantum force 2}) and (\ref{intermediate result for quantum force 4}), obtaining: 
\begin{align} 
\vec f^\textnormal{A}_{\textnormal{qu}}(\vec q,t) =& - \nabla_{\vec q} \left( \hspace{0.05 cm} \underline{\underline{1}} \hspace{0.05 cm}  P_\textnormal{A} \right) -  N(\textnormal{A}) m_\textnormal{A} \sum_{\textnormal{B},i} 
\hspace{-0.1 cm} \int \hspace{-0.1 cm} d \vec Q \hspace{0.1 cm} \delta(\vec q - \vec q_1^{\hspace{0.05 cm} \textnormal{A}} )  \hspace{0.1 cm} 
\nabla_i^\textnormal{B} \hspace{-0.05 cm} \left[ D \left( \vec d_{i}^{\hspace{0.07 cm}\textnormal{B}} \otimes \vec d_{1}^{\hspace{0.05cm} \textnormal{A}}  \right) \right]. \label{intermediate result for quantum force 5}
\end{align}
Now, we again use a similar argumentation to the transformation of Eqn.\ (\ref{right_side_transformed_MPBCE}) to Eqn.\ (\ref{Preresult_right_for_MPQHDCE}) that in Eqn.\ (\ref{intermediate result for quantum force 5}) only the summand for $\{\textnormal{B}=\textnormal{A}, i=1\}$ does not vanish. Finally, using the definition (\ref{quantum_momentum_flow_densitytensor}) for the quantum part $\underline{\underline{\Pi}}^\textnormal{A,qu}(\vec q,t)$ of the momentum flow density tensor for the particle of the sort \textnormal{A}, Eqn.\ (\ref{Connection_quantum_force_force}) can be proven as: 
\begin{align}
\vec f^\textnormal{A}_{\textnormal{qu}}(\vec q,t) &= - \nabla_{\vec q} \left( \hspace{0.05 cm} \underline{\underline{1}} \hspace{0.05 cm}  P_\textnormal{A} \right) -  \nabla_{\vec q} \left[ N(\textnormal{A}) m_\textnormal{A}      
 \int d \vec Q \hspace{0.1 cm} \delta(\vec q - \vec q_1^{\hspace{0.05 cm} \textnormal{A}} )  \hspace{0.1 cm} 
 D \left( \vec d_{1}^{\hspace{0.07 cm}\textnormal{A}} \otimes \vec d_{1}^{\hspace{0.05cm} \textnormal{A}} \right) \right] \nonumber \\
 & =  - \nabla_{\vec q} \hspace{0.05cm} \underline{\underline{\Pi}}^\textnormal{A,qu}(\vec q,t). \label{intermediate result for quantum force 6}
\end{align}
Combining Eqns.\ (\ref{MPBEM_transformed_rewritten}) and (\ref{Connection_quantum_force_force}), we now find for the right side of Eqn.\ (\ref{MPBEM_transformed})
the following result:
\begin{align}
- N(\textnormal{A})  \hspace{-0.1 cm} \int \hspace{-0.1 cm} d \vec Q \hspace{0.1 cm} \delta(\vec q - \vec q_1^{\hspace{0.05 cm} \textnormal{A}})  \hspace{0.1 cm}  D(\vec Q,t) \hspace{0.1 cm} \nabla_1^\textnormal{A} \left [ V_{\textnormal{qu}}(\vec Q,t) + V(\vec Q,t) \right] =& \; \vec f^\textnormal{A}(\vec q,t) - \nabla_{\vec q} \hspace{0.05 cm}  \underline{\underline{\Pi}}^{\textnormal{A,qu}}(\vec q,t).  \label{final result for quantum force 6}
\end{align}
As the next task, we transform the left side of Eqn.\ (\ref{MPBEM_transformed}) by analyzing the $x$ component of this equation in detail: 
\begin{align}
     &\left \lbrace N(\textnormal{A}) m_\textnormal{A} \hspace{-0.1 cm} \int \hspace{-0.1 cm} d \vec Q \hspace{0.1 cm} \delta(\vec q - \vec q_1^{\hspace{0.05 cm} \textnormal{A}} )  \hspace{0.05 cm}  D \hspace{-0.05 cm}  \left[\frac{\partial}{\partial t} + \sum_{B,i} \left( \vec w_i^\textnormal{B} \nabla_i^\textnormal{B} \right) \right] \vec w_1^\textnormal{A}         
     \right \rbrace_x = \nonumber \\
= \; &  N(\textnormal{A}) m_\textnormal{A} \hspace{-0.1 cm} \int \hspace{-0.1 cm} d \vec Q \hspace{0.1 cm} \delta(\vec q - \vec q_1^{\hspace{0.05 cm} \textnormal{A}} )  \hspace{0.05 cm}  D \hspace{-0.05 cm}   \left [ \frac{\partial }{\partial t} +  \sum_{\textnormal{B},i} \left( \vec w_i^\textnormal{B} \nabla_i^\textnormal{B} \right) \right]  \hspace{-0.07 cm}   w_{1x}^\textnormal{A} \; = \nonumber \\
= \; &  N(\textnormal{A}) m_\textnormal{A} \hspace{-0.1 cm} \int \hspace{-0.1 cm} d \vec Q \hspace{0.1 cm} \delta(\vec q - \vec q_1^{\hspace{0.05 cm} \textnormal{A}} )   \left[ \frac{\partial}{\partial t} \left( D w_{1x}^\textnormal{A} \right) -  w_{1x}^\textnormal{A} \frac{\partial D}{\partial t}\right] \nonumber \\
     & + \; N(\textnormal{A}) m_\textnormal{A} \hspace{-0.1 cm}  \int \hspace{-0.1 cm} d \vec Q \hspace{0.1 cm} \delta(\vec q - \vec q_1^{\hspace{0.05 cm} \textnormal{A}} )  \hspace{0.05 cm}  \hspace{0.05 cm}  D  
     \left \lbrace  \sum_{\textnormal{B},i} \left[ \sum_\alpha  \frac{\partial}{\partial q_{i \alpha}^{\textnormal{B}}} \left( w_{1 x}^{\textnormal{A}} w_{i \alpha}^{\textnormal{B}}  \right)   \right] - w_{1 x}^{\textnormal{A}} \nabla_i^{\textnormal{B}} \vec w_i^{\textnormal{B}}    \right \rbrace \nonumber 
\end{align} \newpage 
\begin{align}     
 = \; & \underbrace{\frac{\partial}{\partial t}   \left[N(\textnormal{A}) m_\textnormal{A} \hspace{-0.1 cm}  \int \hspace{-0.1 cm} d \vec Q \hspace{0.1 cm} \delta(\vec q - \vec q_1^{\hspace{0.05 cm} \textnormal{A}} )  \hspace{0.05 cm}  \hspace{0.05 cm}  D  \hspace{0.1 cm} w_{1 x}^\textnormal{A} \right] }_{ \overset{\text{(\ref{mass current density})}}{\underset{\text{}}{=}} \; \frac{\partial}{\partial t} j_{m,x}^\textnormal{A}} \textnormal{\hspace{5.0 cm}} \nonumber \\
& - \; N(\textnormal{A}) m_\textnormal{A} \hspace{-0.1 cm}  \int \hspace{-0.1 cm} d \vec Q \hspace{0.1 cm} \delta(\vec q - \vec q_1^{\hspace{0.05 cm} \textnormal{A}} )  \hspace{0.05 cm}  \hspace{0.05 cm} w_{1 x}^\textnormal{A}  \hspace{-1.0 cm} \underbrace{\frac{\partial D}{\partial t}}_{\overset{\text{(\ref{first_version_MPBCE})}}{\underset{\text{}}{=}} \; - \sum_{\textnormal{B},i} \hspace{-0.05 cm} \nabla_i^\textnormal{B} \left( D \vec w_i^\textnormal{B} \right)}  \nonumber \\
& + \; \underbrace{N(\textnormal{A}) m_\textnormal{A} \hspace{-0.1 cm}  \int \hspace{-0.1 cm} d \vec Q \hspace{0.1 cm} \delta(\vec q - \vec q_1^{\hspace{0.05 cm} \textnormal{A}}) \hspace{0.05 cm} \sum_{\textnormal{B},i} \left[ \sum_{\alpha} \frac{\partial}{\partial q_{i \alpha}^\textnormal{B}} \left( D \hspace{0.05 cm} w_{1 x}^\textnormal{A}  w_{i \alpha}^\textnormal{B} \right) \right]}_{= \;  \sum_{\alpha} \frac{\partial}{\partial q_\alpha} \left[ N(\textnormal{A}) m_\textnormal{A} \hspace{-0.05 cm} \int  \hspace{-0.05 cm}  d \vec Q \hspace{0.1 cm}   \delta(\vec q - \vec q_1^{\hspace{0.05 cm} \textnormal{A}}) \hspace{0.05 cm} D \hspace{0.05 cm} w_{1 x}^\textnormal{A}  w_{1 \alpha}^\textnormal{A}  \right]} \nonumber \\
& - \; N(\textnormal{A}) m_\textnormal{A} \hspace{-0.1 cm}  \int \hspace{-0.1 cm} d \vec Q \hspace{0.1 cm} \delta(\vec q - \vec q_1^{\hspace{0.05 cm} \textnormal{A}}) \hspace{0.05 cm}  w_{1 x}^\textnormal{A} \sum_{\textnormal{B},i} \underbrace{\left( \sum_{\alpha} w_{i \alpha}^\textnormal{B}  \frac{\partial D}{\partial q_{i \alpha}^\textnormal{B} } \right)}_{= \; \vec w_{i}^\textnormal{B} \nabla_i^\textnormal{B} D}  \nonumber \\ 
& - \; N(\textnormal{A}) m_\textnormal{A} \hspace{-0.1 cm}  \int \hspace{-0.1 cm} d \vec Q \hspace{0.1 cm} \delta(\vec q - \vec q_1^{\hspace{0.05 cm} \textnormal{A}}) \hspace{0.05 cm} D \hspace{0.05cm} w_{1 x}^\textnormal{A} \sum_{\textnormal{B},i} \nabla_i^\textnormal{B} \vec w_i^\textnormal{B}. \label{MPBEM_transformed_left_2}
\end{align}
So, we got an intermediate result (\ref{MPBEM_transformed_left_2}) for this vector component that ranges over five equation lines. As indicated by lowered braces, the first of these five equation lines can be transformed using the $x$ component of the vector equation (\ref{mass current density}) for the MPQHD mass current density $\vec j^\textnormal{A}_m(\vec q,t)$, the second equation line can be transformed using the continuity equation of BM (\ref{first_version_MPBCE}), the third equation line can be simplified by regarding the fact that due to the divergence theorem only the summand for $\{\textnormal{B}=\textnormal{A},i=1\}$ in the double sum over $\textnormal{B}$ and $i$ does not vanish, and the fourth equation line is rewritten by using a vector notation instead of the sum over $\alpha$. In addition, we permute the second and the third of these five equation lines and find: \newpage
\begin{align}
& \left \lbrace N(\textnormal{A}) m_\textnormal{A} \hspace{-0.1 cm} \int \hspace{-0.1 cm} d \vec Q \hspace{0.1 cm} \delta(\vec q - \vec q_1^{\hspace{0.05 cm} \textnormal{A}} )  \hspace{0.1 cm}  D   \left[ \frac{\partial}{\partial t} + \sum_{\textnormal{B},i} \left( \vec w_i^\textnormal{B} \nabla_i^\textnormal{B} \right) \right] \hspace{-0.1 cm}  \vec w_1^\textnormal{A} \right \rbrace_x =  \nonumber \\
= & \; \frac{\partial j_{m,x}^\textnormal{A}}{\partial t} \nonumber \\
& + \; \underbrace{\sum_{\alpha} \frac{\partial}{\partial q_\alpha} \left[ N(\textnormal{A}) m_\textnormal{A} \hspace{-0.05 cm} \int  \hspace{-0.05 cm}  d \vec Q \hspace{0.1 cm}   \delta(\vec q - \vec q_1^{\hspace{0.05 cm} \textnormal{A}}) \hspace{0.05 cm} D \hspace{0.05 cm} w_{1 x}^\textnormal{A}  w_{1 \alpha}^\textnormal{A}  \right]}_{
\overset{\text{(\ref{classic_momentum_flow_densitytensor})}}{\underset{\text{}}{=}}
\; \left( \nabla_{\vec q} \hspace{0.03 cm} \underline{\underline{\Pi}}^\textnormal{A,cl} \right)_x} \nonumber \\
& + \; N(\textnormal{A}) m_\textnormal{A} \hspace{-0.1 cm}  \int \hspace{-0.1 cm} d \vec Q \hspace{0.1 cm} \delta(\vec q - \vec q_1^{\hspace{0.05 cm} \textnormal{A}} )  \hspace{0.05 cm}  \hspace{0.05 cm} w_{1 x}^\textnormal{A} \sum_{\textnormal{B},i} \hspace{-0.05 cm} \nabla_i^\textnormal{B} \left( D \hspace{0.05 cm} \vec w_i^\textnormal{B} \right) \nonumber \\
& - \; N(\textnormal{A}) m_\textnormal{A} \hspace{-0.1 cm}  \int \hspace{-0.1 cm} d \vec Q \hspace{0.1 cm} \delta(\vec q - \vec q_1^{\hspace{0.05 cm} \textnormal{A}}) \hspace{0.05 cm}  w_{1 x}^\textnormal{A}  \sum_{\textnormal{B},i} \vec w_{i}^\textnormal{B} \hspace{0.05 cm} \nabla_i^\textnormal{B} D  
\nonumber \\ 
& - \; N(\textnormal{A}) m_\textnormal{A} \hspace{-0.1 cm}  \int \hspace{-0.1 cm} d \vec Q \hspace{0.1 cm} \delta(\vec q - \vec q_1^{\hspace{0.05 cm} \textnormal{A}}) \hspace{0.05 cm} D \hspace{0.05cm} w_{1 x}^\textnormal{A} \sum_{\textnormal{B},i} \nabla_i^\textnormal{B} \vec w_i^\textnormal{B}. \label{MPBEM_transformed_left_3}
\end{align}
Above we indicated, using within a lowered brace Eqn.\ (\ref{classic_momentum_flow_densitytensor}), that one finds the term appearing in the second line of the right side of Eqn.\ (\ref{MPBEM_transformed_left_3}) to be equal to the $x$ component of the divergence of classical part $\underline{\underline{\Pi}}^{\textnormal{A,cl}}(\vec q,t)$ of the momentum flow density tensor for the particles of the sort $\textnormal{A}$. Moreover, we merge the terms appearing in the third, fourth, and fifth line of the right side of Eqn.\ (\ref{MPBEM_transformed_left_3}) and recognize that the sum of these terms vanishes. Thus, we get the following result for the $x$ component of the expression on the left side of Eqn.\ (\ref{MPBEM_transformed}):
\begin{align}
& \left \lbrace N(\textnormal{A}) m_\textnormal{A} \hspace{-0.1 cm} \int \hspace{-0.1 cm} d \vec Q \hspace{0.1 cm} \delta(\vec q - \vec q_1^{\hspace{0.05 cm} \textnormal{A}} )  \hspace{0.1 cm}  D   \left[ \frac{\partial}{\partial t} + \sum_{\textnormal{B},i} \left( \vec w_i^\textnormal{B} \nabla_i^\textnormal{B} \right) \right] \hspace{-0.1 cm}  \vec w_1^\textnormal{A} \right \rbrace_x =  \nonumber \\
 = & \; \frac{\partial j_{m,x}^\textnormal{A}}{\partial t}  +  \left( \nabla_{\vec q} \hspace{0.03 cm} \underline{\underline{\Pi}}^\textnormal{A,cl} \right)_x \nonumber \\
& + \; N(\textnormal{A}) m_\textnormal{A} \hspace{-0.1 cm}  \int \hspace{-0.1 cm} d \vec Q \hspace{0.1 cm} \delta(\vec q - \vec q_1^{\hspace{0.05 cm} \textnormal{A}} )  \hspace{0.05 cm}  \hspace{0.05 cm} w_{1 x}^\textnormal{A} \sum_{\textnormal{B},i} \hspace{-0.05 cm} \underbrace{\left[ \nabla_i^\textnormal{B} \left( D \hspace{0.05 cm} \vec w_i^\textnormal{B} \right) - \vec w_i^\textnormal{B}  \hspace{+0.03 cm}  \nabla_i^\textnormal{B} D  - D  \hspace{+0.03 cm} \nabla_i^\textnormal{B} \vec w_i^\textnormal{B} \right]}_{= \; 0} \nonumber \\
= &  \; \frac{\partial j_{m,x}^\textnormal{A}}{\partial t}  +  \left( \nabla_{\vec q} \hspace{0.03 cm} \underline{\underline{\Pi}}^\textnormal{A,cl} \right)_x.
\end{align}
Therefore, the expression on the left side of Eqn.\ (\ref{MPBEM_transformed}) is given by: 
\begin{align}
N(\textnormal{A}) m_\textnormal{A} \hspace{-0.1 cm} \int \hspace{-0.1 cm} d \vec Q \hspace{0.1 cm} \delta(\vec q - \vec q_1^{\hspace{0.05 cm} \textnormal{A}} )  \hspace{0.1 cm}  D   \left[ \frac{\partial}{\partial t} + \sum_{\textnormal{B},i} \left( \vec w_i^\textnormal{B} \nabla_i^\textnormal{B} \right) \right] \hspace{-0.1 cm}  \vec w_1^\textnormal{A} 
& = \frac{\partial \vec j_{m}^\textnormal{A} (\vec q,t)}{\partial t}  +  \nabla_{\vec q} \hspace{0.03 cm} \underline{\underline{\Pi}}^\textnormal{A,cl}(\vec q,t). \label{MPBEM_transformed_left_4}
\end{align}
Now, we combine the result (\ref{MPBEM_transformed_left_4}) for the left side of Eqn.\ (\ref{MPBEM_transformed}) and the result (\ref{final result for quantum force 6}) for the right side of Eqn.\ (\ref{MPBEM_transformed}) and get: 
\begin{align}
\frac{\partial \vec j_{m}^\textnormal{A} (\vec q,t)}{\partial t}  +  \nabla_{\vec q} \hspace{0.03 cm} \underline{\underline{\Pi}}^\textnormal{A,cl}(\vec q,t) =& \; \vec f^\textnormal{A}(\vec q,t) - \nabla_{\vec q} \hspace{0.05 cm}  \underline{\underline{\Pi}}^{\textnormal{A,qu}}(\vec q,t).   \label{nearly_MPEEM}
\end{align}
By shifting the term $\nabla_{\vec q} \hspace{0.03 cm} \underline{\underline{\Pi}}^\textnormal{A,cl}(\vec q,t)$ to the right side of Eqn.\ (\ref{nearly_MPEEM}) and 
by applying $\underline{\underline{\Pi}}^{\textnormal{A}}(\vec q,t) = \underline{\underline{\Pi}}^{\textnormal{A,cl}}(\vec q,t) + \underline{\underline{\Pi}}^{\textnormal{A,qu}}(\vec q,t)$, one gets the Ehrenfest equation of motion for the particles of the sort $\textnormal{A}$ that we presented already in \cite{Renziehausen_2017}: 
\begin{align}
\frac{\partial \vec j_{m}^\textnormal{A} (\vec q,t)}{\partial t}  =& \; \vec f^\textnormal{A}(\vec q,t) - \nabla_{\vec q} \hspace{0.05 cm}  \underline{\underline{\Pi}}^{\textnormal{A}}(\vec q,t). \label{again_MPEEM}
\end{align}
In addition, in \cite{Renziehausen_2017}, it was shown that performing some mathematical operations using the continuity equation of MPQHD for the particles of the sort $\textnormal{A}$ (\ref{MPQHDCE}), one can transform the Ehrenfest equation of motion for the particles of the sort $\textnormal{A}$ (\ref{again_MPEEM}) into the quantum Cauchy equation for the particles of this sort, which is given by: 
\begin{align}
\rho^\textnormal{A}_m(\vec q,t) \left[ \frac{\partial}{\partial t} + \left(  \vec v^\textnormal{A}(\vec q,t) \nabla_{\vec q} \right) \right] \vec v^\textnormal{A}(\vec q,t)  &= \vec f^{\hspace{0.05 cm \textnormal{A}}}(\vec q,t)  - \nabla_{\vec q} \hspace{0.05 cm} \underline{\underline{p}}^\textnormal{A}(\vec q,t). \label{MPCEM_sort_A}
\end{align}
In the quantum Cauchy equation for the particles of the sort $\textnormal{A}$ the MPQHD particle velo\-city $\vec v^\textnormal{A}(\vec q,t)$ for the particles of the sort $\textnormal{A}$ and the pressure tensor $\underline{\underline{p}}^\textnormal{A}(\vec q,t)$ for the particles of this sort appear. The definition of these quantities was discussed already in our prework \cite{Renziehausen_2017}. For the following discussions, there is no need to recapitulate these definitions again, thus, we skip them in this work. \newline
Taking into account that the derivation mentioned above, i.\ e. the derivation of the quantum Cauchy equation for the particles of the sort $\textnormal{A}$ (\ref{MPCEM_sort_A}),  was already worked out in detail in our previous work, we have proven that this quantum hydrodynamic equation can be derived from the Eulerian version of the Bohmian equation of motion (\ref{Eulerian_MPBEM}) as a starting point. \newline \newline 
Now, the only remaining open task of the outlined tasks at the end of Sec.\ \ref{Basic Physics} is the proof that the quantum Cauchy equation for the total particle ensemble can be derived from the Eulerian version of the Bohmian equation of motion (\ref{Eulerian_MPBEM}) as a starting point. Therefore, we regrad as shown above that Eqn.\ (\ref{Eulerian_MPBEM}) can be used as a starting point to derive the Ehrenfest equation of motion for the particles of the sort $\textnormal{A}$ (\ref{again_MPEEM}). \newline 
As the next step, we here recapitulate shortly that we have proven already in our prework \cite{Renziehausen_2017} that one can derive the quantum Cauchy equation for the total particle ensemble using the Ehrenfest equation of motion for the particles of the sort $\textnormal{A}$ (\ref{again_MPEEM}) and the continuity equation of MPQHD for the total particle ensemble (\ref{MPQHDCE_total}): \newline 
Therefore, we take into account that, analogously to the definition (\ref{def_j_tot}) of the MPQHD total mass current density $\vec j^{\hspace{0.05 cm} \textnormal{tot}}_m(\vec q,t)$, the momentum flow density tensor $\underline{\underline{\Pi}}^\textnormal{tot}(\vec q,t)$ and the force density $\vec f^{\hspace{0.05 cm}\textnormal{tot}}(\vec q,t)$ for the total particle ensemble are given by the sum over the correspondent quantities for the particular particle sorts: 
\begin{align}
\underline{\underline{\Pi}}^\textnormal{tot}(\vec q,t) &= \sum_{\textnormal{A}=1}^{N_S} \underline{\underline{\Pi}}^\textnormal{A}(\vec q,t), \label{def_Pi_tot}  \\
\vec f^{\hspace{0.05 cm} \textnormal{tot}}(\vec q,t) &= \sum_{\textnormal{A}=1}^{N_S} \vec f^\textnormal{A}(\vec q,t). \label{def_f_tot}
\end{align}
Now, we can sum up the Ehrenfest equation of motion for a certain sort of particle $\textnormal{A}$ (\ref{again_MPEEM}) over all sorts of particle and using the Eqns.\ (\ref{def_j_tot}), (\ref{def_Pi_tot}), and (\ref{def_f_tot}), one gets the Ehrenfest equation of motion for the total particle ensemble as a result. This equation is given by: 
\begin{align}
\frac{\partial \vec j_{m}^{\hspace{0.05 cm} \textnormal{tot}} (\vec q,t)}{\partial t}  =& \; \vec f^\textnormal{\hspace{0.05 cm}\textnormal{tot}}(\vec q,t) - \nabla_{\vec q} \hspace{0.05 cm}  \underline{\underline{\Pi}}^{\hspace{0.05 cm} \textnormal{tot}}(\vec q,t). \label{again_MPEEM_tot}
\end{align}
In addition, in a calculation using the continuity equation of MPQHD for the total particle ensemble (\ref{MPQHDCE_total}), one can now transform the Ehrenfest equation of motion for the total particle ensemble (\ref{again_MPEEM_tot}) into the quantum Cauchy equation for the total particle ensemble. Here, we skip the details of this calculation and only state its result, the quantum Cauchy equation for the total particle ensemble \cite{Renziehausen_2017}: 
\begin{align}
\rho^\textnormal{tot}_m(\vec q,t) \left[ \frac{\partial}{\partial t} + \left(  \vec v^\textnormal{\hspace{0.05 cm}tot}(\vec q,t) \nabla_{\vec q} \right) \right] \vec v^\textnormal{\hspace{0.05 cm}tot}(\vec q,t) 
&= \vec f^{\hspace{0.05 cm \textnormal{tot}}}(\vec q,t)  - \nabla_{\vec q} \hspace{0.05 cm} \underline{\underline{p}}^\textnormal{tot}(\vec q,t). \label{MPCEM_tot}
\end{align}
In Eqn.\ (\ref{MPCEM_tot}), the quantity $\vec v^{\hspace{0.05 cm} \textnormal{tot}}(\vec q,t)$ is the MPQHD particle velocity for the total particle ensemble, 
and $\underline{\underline{p}}^\textnormal{tot}(\vec q,t)$ is the pressure tensor for the total particle ensemble. Like for the corresponding quantities for a single sort of particle $\textnormal{A}$, we refer for the definitions of the quantities $\vec v^{\hspace{0.05 cm} \textnormal{tot}}(\vec q,t)$ and  $\underline{\underline{p}}^\textnormal{tot}(\vec q,t)$ to our previous work \cite{Renziehausen_2017}. \newline 
So, we now summarized how the quantum Cauchy equation for the total particle ensemble (\ref{MPCEM_tot}) can be derived using the Eulerian version of the Bohmian equation of motion (\ref{Eulerian_MPBEM}) as a starting point. \newline \newline 
As an additional remark to this proof, we mention that the reader might wonder why this proof was not performed by trying to find the quantum Cauchy equation for the total particle ensemble (\ref{MPCEM_tot}) by summing up the quantum Cauchy equation for a certain sort of particle $\textnormal{A}$  (\ref{MPCEM_sort_A}) over all sorts of particle instead of summing up the Ehrenfest equation of motion for a certain sort of particle \textnormal{A} (\ref{again_MPEEM}) over all sorts of particle. 
\newline \noindent The reason why we did not choose this proceeding in our analysis above is that one does not get the quantum Cauchy equation for the total particle ensemble (\ref{MPCEM_tot}) just by summing up the quantum Cauchy equation for a certain sort of particle $\textnormal{A}$ (\ref{MPCEM_sort_A}) over all sorts of particle because the Eqn.\ (\ref{MPCEM_sort_A}) is a non-linear differential equation where the superposition principle is not valid. 
\newline \newline \noindent For a good overview of the equations of BM, which we used as a starting point in this work, and the equations of MPQHD, which we here found as final results, we present them again in Tab. \ref{Table2}. 
\begin{table} [t!]  
\begin{tabular}{|c|c|} \hline
 {\small Bohmian equations}                                                        & {\small MPQHD equations}                               \\ \hline \hline
                                                                                                                          &               \\
                                                                                                                          &  $\frac{\partial}{\partial t} \rho_m^\textnormal{A}(\vec q,t) = - \nabla_{\vec q} \hspace{0.05 cm} \vec j_m^\textnormal{A}(\vec q,t)$              \\
                                                                                                                          &   continuity equation of MPQHD \\ 
                                                                                                                          &  for the particle sort $\textnormal{A}$ \\
 $\frac{\partial}{\partial t} D(\vec Q,t) = -\nabla_{\vec Q} \vec J(\vec Q,t)$                                            &  \\                                                                                                                                      continuity equation of BM &                               \\
                                                                                                                          &  $\frac{\partial}{\partial t} \rho_m^\textnormal{tot}(\vec q,t) = - \nabla_{\vec q} \hspace{0.05 cm} \vec j_m^\textnormal{tot}(\vec q,t) $          \\ 
                                                                                                                          & continuity equation of MPQHD \\
                                                                                                                          & for the total particle ensemble   \\ & \\ \hline  
                                                                                                                          &               \\
                                                                                                                          & $\rho_m^\textnormal{A}(\vec q,t) \left[ \frac{\partial}{\partial t} + \left(  \vec v^\textnormal{A}(\vec q,t) \nabla_{\vec q} \right) \right] \vec v^\textnormal{A}(\vec q,t)  =$ \\
                                                                                                                          & $ = \vec f^\textnormal{A}(\vec q,t)  - \nabla_{\vec q} \hspace{0.05 cm} \underline{\underline{p}}^\textnormal{A}(\vec q,t)$            \\
                                                                                                                          &      quantum Cauchy equation  \\ 
                                                                                                                          &      for the particle sort $\textnormal{A}$ \\
 $\left[\frac{\partial}{\partial t} + \sum_{\textnormal{A},i} \vec w_i^\textnormal{A}(\vec Q,t) 
 \nabla_i^\textnormal{A} \right] \vec p(\vec Q,t) =$                                                                      &                                             \\
  $= -\nabla_{\vec Q} [  V(\vec Q,t) + V_{\textnormal{qu}}(\vec Q,t) ]$                                                   &                                             \\    
 Eulerian version of the                                                                                                        &                                             \\                                              
 Bohmian equation of motion                                                                             &                                             \\
                                                                                                                          &  $\rho_m^\textnormal{tot}(\vec q,t)   \left[ \frac{\partial}{\partial t} + \left(  \vec v^\textnormal{\hspace{0.05 cm}tot}(\vec q,t) \nabla_{\vec q} \right) \right] \vec v^\textnormal{\hspace{0.05 cm}tot}(\vec q,t)  =$\\ 
                                                                                                                          & $= \vec f^{\hspace{0.05 cm}\textnormal{tot}}(\vec q,t) - \nabla_{\vec q} \hspace{0.05 cm} \underline{\underline{p}}^\textnormal{tot}(\vec q,t)$            \\ 
                                                                                                                          &   quantum Cauchy equation \\
                                                                                                                          &   for the total particle ensemble       \\ & \\ \hline  
\end{tabular}
\caption{\noindent On the left side of this table, we present the Bohmian equations, where all quantities depend on the complete set of particle coordinates $\vec Q$ and the time $t$, and on the right side the corresponding quantum hydrodynamic equations, where all quantities depend on a single postition vector $\vec q$ and the time $t$.} 
\label{Table2}
\end{table} 
\newline \newline    
As an additional point, we mention this analogy between classical mechanics and quantum mechanics: \newline In a classic system with several interacting particles, sol\-ving the coupled Newtonian equations of motion to receive the trajectories of these particles is a promising strategy for small systems with only few particles because already for relative large systems with tens or hundreds of particles the system of the Newtonian equations of motion is very hard to solve due to its big size. Thus, for such large many-particle systems, using classical hydrodynamics instead is a better approach because for these systems, the size of the hydrodynamical differential equations is much smaller than the size of the system of the Newtonain differential equations. \newline 
In an analogous manner, solving the Bohmian equation of motion (\ref{Lagrangian_MPBEM}) for a quantum system with several interacting particles to calculate the set $\vec p(t)$ of all the individual Bohmian particle momenta is also a promising strategy for small systems only. The reason why solving the Bohmian equation of motion (\ref{Lagrangian_MPBEM}) is a good ansatz for small systems only is that already for relative large many-particle systems with tens or hundreds of particles the Bohmian equation of motion (\ref{Lagrangian_MPBEM}) is difficult to solve since for these systems, the Bohmian equation of motion (\ref{Lagrangian_MPBEM}) is a high-dimensional diffe\-rential equation with $[\sum_{\textnormal{A}=1}^{N_S} 3 N(\textnormal{A})]$ dimensions. As a contrast, for many-particle systems, solving the Ehrenfest equations of motion (\ref{again_MPEEM}) and (\ref{again_MPEEM_tot}) for a calculation of the MPQHD mass current densities $\vec j_m^\textnormal{A}(\vec q,t)$ and $\vec j_m^\textnormal{tot}(\vec q,t)$, or sol\-ving the quantum Cauchy equations (\ref{MPCEM_sort_A}) and (\ref{MPCEM_tot}) for a calculation of the MPQHD particle velo\-cities $\vec v^\textnormal{A}(\vec q,t)$ and $\vec v^{\hspace{0.05 cm} \textnormal{tot}}(\vec q,t)$, are more promising approaches than solving the Bohmian equation of motion (\ref{MPCEM_sort_A}) because the Ehrenfest equations of motion and quantum Cauchy equations (\ref{again_MPEEM}), (\ref{MPCEM_sort_A}), (\ref{again_MPEEM_tot}), and (\ref{MPCEM_tot}) are always three-dimensional equations independent of the size of the system. \newline 
Now, the reader might object that applying MPQHD for the calculation of the current densities $\vec j_m^\textnormal{A}(\vec q,t)$, $\vec j_m^\textnormal{tot}(\vec q,t)$ using Eqns.\ (\ref{again_MPEEM}) and  (\ref{again_MPEEM_tot}) is not purposive because in order to apply MPQHD in the manner described in the previous paragraph, the wave function of the system $\Psi(\vec Q,t)$ has to be available in order to know the quantities on the right side of the Eqns. (\ref{again_MPEEM}) and  (\ref{again_MPEEM_tot})  -- and knowing the wave function $\Psi(\vec Q,t)$ we could calculate the mass current densities $\vec j_m^\textnormal{A}(\vec q,t)$ and $\vec j_m^\textnormal{tot}(\vec q,t)$ directly, anyway: Hereby, this direct calculation would be done using the Eqns.\ (\ref{Bohm representation}), (\ref{Def_total_particle_density}), (\ref{Def_velocity}), (\ref{Def_BM_current_density}), (\ref{mass current density}),  and (\ref{def_j_tot}). 
However, there are still systems where the application of MPQHD is interesting: \newline 
In vibrating molecules within a single electronic state, electronic mass current densities $\vec j^e_m(\vec q,t)$ vanish if one calculates them directly within the Born-Oppenheimer approximation (BOA) using the BOA wave function 
\begin{align}
\Psi_{\textnormal{BOA}}(\vec Q^e,\vec Q^n,t)  &= a_{\textnormal{BOA}}(\vec Q^e,\vec Q^n,t) \exp \left[ \frac{\mathrm{i} \hspace{0.05 cm} S_{\textnormal{BOA}}(\vec Q^n,t)}{\hbar} \right]  
\end{align}
as an input \cite{Barth_2009}, where $\vec Q^e$ is the complete set of all electronic coordinates and $\vec Q^n$ is the complete set of all nuclear coordinates. The reason for this is that the action $S_{\textnormal{BOA}}(\vec Q^n,t)$ for the occupation of a single electronic state is falsely independent of the set of electronic coordinates $\vec Q^e$ because of the BOA. So, due to Eqn.\ (\ref{Def_velocity}), all the Bohmian electronic velo\-cities $\vec w_i^e(\vec Q,t)$, $i \in \{1,2,\ldots,N_e\}$ for all electrons vanish.  \newline
As a contrast, the total particle density $D(\vec Q^e,\vec Q^n,t)$ can be calculated within the BOA in good accuracy using
\begin{align}
D(\vec Q^e, \vec Q^n,t) \approx a_{\textnormal{BOA}}^2(\vec Q^e, \vec Q^n,t).
\end{align}
In addition, if one analyzes molecular vibrations where electrons and nuclei are at rest at the start time $t=0$, one can assume as a start condition the electronic mass current density $\vec j_m^e(\vec q,t)$ to be zero at this start time. \newline 
Now, we discuss the application of the Ehrenfest equation of motion (\ref{again_MPEEM}) for the particle sort $\textnormal{A}$ for this system: Therefore, regarding the electrons as the particle sort $\textnormal{A}$ in Eqn.\ (\ref{again_MPEEM}), we take into account that on the right side of this equation the force density $\vec f^e(\vec q,t)$ and the electronic momentum flow density tensor $\underline{\underline{\Pi}}^e(\vec q,t)$ appear. Hereby, due to Eqn.\ (\ref{split_tensor}), the tensor $\underline{\underline{\Pi}}^e(\vec q,t)$ can be split into a classical part $\underline{\underline{\Pi}}^{e,\textnormal{cl}}(\vec q,t)$ and a quantum part $\underline{\underline{\Pi}}^{e,\textnormal{qu}}(\vec q,t)$. As the next step, we realize that the BOA wave function $\Psi_{\textnormal{BOA}}(\vec Q^e,\vec Q^n,t)$ has on the quantities $\vec f^e(\vec q,t)$ and $\underline{\underline{\Pi}}^{e,\textnormal{qu}}(\vec q,t)$ only an impact via the total particle density $D(\vec Q^e,\vec Q^n,t)$; this can be checked using the Eqns.\ (\ref{force_density}), (\ref{scalar_pressure}), (\ref{definition d}), and (\ref{quantum_momentum_flow_densitytensor}). So, we can calculate these quantities $\vec f^e(\vec q,t)$ and $\underline{\underline{\Pi}}^{e,\textnormal{qu}}(\vec q,t)$ within the BOA in good accuracy. However, the BOA wave function $\Psi_{\textnormal{BOA}}(\vec Q^e,\vec Q^n,t)$ has on the classical tensor $\underline{\underline{\Pi}}^{e,\textnormal{cl}}(\vec q,t)$ both an impact via the total particle density $D(\vec Q^e,\vec Q^n,t)$ and via the electronic Bohmian velocity $\vec w_1^e(\vec Q^e,\vec Q^n,t)$, which vanishes untruely in the BOA. As a consequence of Eqn.\ (\ref{momentum_flow_densitytensor no N_k tensor version}), this tensor $\underline{\underline{\Pi}}^{e,\textnormal{cl}}(\vec q,t)$ vanishes untruely in the BOA, too. However, for situations where the term $-\nabla \underline{\underline{\Pi}}^{e,\textnormal{cl}}(\vec q,t)$ is negligble, anyway, compared to the term $[\vec f^{e}(\vec q,t) - \nabla \underline{\underline{\Pi}}^{e,\textnormal{qu}}(\vec q,t)]$,  the right side of the Ehrenfest equation of motion (\ref{again_MPEEM}) can be calculated within the BOA in good approximation. Thus, for molecular vibrations where electrons and nuclei are at rest at the start time $t=0$, this condition is fulfilled at least at small times $t$ after the start time. So, at these times, we can calculate the right side of Eqn. (\ref{again_MPEEM}) in the BOA in good accuracy and are able to solve this differential equation numerically using the start condition $\vec j_m^e(\vec q,t=0) = \vec 0$ mentioned above. Further research will now be necessary to assess the extent to which this approach can be used to investigate vibrations in molecules.   
\section{Summary} \label{Summary}
In the early 1950s \cite{Bohm_1952_1, Bohm_1952_2}, Bohm used an ansatz to transform the Schrödinger equation which leads to the result that the Schrödinger equation can be split in two other differential equations: The first of these equations is the continuity equation of BM that is related to the conservation of particles, and the second of these equations is the Bohmian equation of motion being very similar to the Newtonian equation of motion except for the detail that in the Bohmian equation of motion an additional quantity appears which is called the quantum potential. These two differential equations are the basis for BM, and they can be derived both for one-particle systems and for many-particle systems, where -- in the general case which we analyze in this work -- the different particles belong to different sorts of particle. 
As an additional point, the Bohmian equation of motion has to be interpreted in the Lagrangian picture, where the observer is following the particles along their trajectories. In the calculations of this work, we use another version of the Bohminan equation of motion in the Eulerian picture, where an observer at a fixed point in space watches the particles move by. It is remarkable that for many-particle systems, the above-mentioned differential equations of BM depend on the complete set of particle coordinates. \newline 
On the other hand, using basic quantum mechanics, Kuzmenkov and Maksimov developed MPQHD in 1999 \cite{Kuzmenkov_1999}, where an averaging over the coordinates of all particles except one is made. Using this ansatz, one can derive the continuity equation of MPQHD, which is a differential equation related to the mass conservation of the system, and two differential equations being related to the momentum conservation of the system, namely the Ehrenfest equation of motion and the quantum Cauchy equation. We picked up this analysis of Kuzmenkov and Maksimov, and in our prework \cite{Renziehausen_2017} we discussed in detail how to extend systematically their discussion for the situation that the system contains different sorts of particle. In this case, one can derive a version of the continuity equation of MPQHD, the Ehrenfest equation of motion, and the quantum Cauchy equation each for the individual particle sorts and for the total particle ensemble. For all these versions of the continuity equation of MPQHD, the Ehrenfest equation of motion and the quantum Cauchy equation, it is valid that they only depend on a single position vector. \newline   
In this work, we analyzed the connection between the above-mentionend differential equations of BM and MPQHD. Using an averaging over the coordinates of all particles except one, we found out how all the versions of the continuity equation of MPQHD and the quantum Cauchy equation can be derived from the continuity equation of BM and the Eulerian version of the Bohmian equation of motion as a starting point -- and during these calculations we got all the versions of the Ehrenfest equation of motion as intermediate results. We think that this article will help other authors to realize how BM and MPQHD are linked. In addition, it can be realized that these fields can be differentiated from each other by the context that the differential equations related to BM depend on the complete set of particle coordinates, but the differential equations related to MPQHD depend on a single position vector only. So, already for relative large many-particle systems consisting of tens or hundreds of particles, the differential equations related to BM are much more complicated to solve than the differential equations related to MPQHD. This is an analogy to the fact that for classical many-particle systems with correspondingly many particles, the Newtonian equations of motion for the trajectories of all the particles of the system are more complicated to solve than the hydrodynamic differential equations. \newline
As a last point, a concrete interesting application for the Ehrenfest equation of motion is the calculation of electronic mass current densities for molecules within one electronic state in the Born-Oppenheimer approximation because for such systems, the direct calculation of these current densities is not possible. Further research is needed to analyze to what extent the MPQHD ansatz is useful to calculate these electronic current densities.


\begin{thebibliography}{99}
\bibitem{Renziehausen_2017_comment_B} In \cite{Renziehausen_2017}, we did not define the Bohmian particle current density $\vec J_i^\textnormal{A}(\vec Q,t)$. So, in this prework, we presented a formula for the calculation of $\vec j_m^\textnormal{A}(\vec q,t)$ where the product $D(\vec Q,t) \; \vec w_i^\textnormal{A}(\vec Q,t)$ instead of the quantity $\vec J_i^\textnormal{A}(\vec Q,t)$ appears.
\end{thebibliography}
\end{document}